\documentclass{article}
\usepackage{siunitx}
\DeclareSIUnit\angstrom{\text{\AA}}
\usepackage{arxiv}
\usepackage[utf8]{inputenc} 
\usepackage[T1]{fontenc}    
\usepackage{hyperref}       
\usepackage{url}            
\usepackage{booktabs}       
\usepackage{amsfonts}       
\usepackage{nicefrac}       
\usepackage{microtype}      
\usepackage{graphicx}
\usepackage{doi}
\usepackage{lipsum}  
\usepackage{graphicx}  
\usepackage{url}       
\usepackage{amsmath}   
\usepackage{siunitx}
\usepackage{url}
\usepackage{braket}
\usepackage{amssymb}
\usepackage{amsmath}
\usepackage{enumerate}
\usepackage{float}
\usepackage{graphicx}
\usepackage{fancyhdr}
\usepackage[style=chem-acs]{biblatex}
\usepackage[most]{tcolorbox}
\usepackage{empheq}
\usepackage{caption}
\newcommand{\tab}{\quad}

\newcommand{\expec}{\mathbb{E}}

\DeclareMathOperator*{\argmax}{arg\,max}

\title{Physics-Informed Gaussian Process Inference of Liquid Structure from Scattering Data}

\author{\href{https://orcid.org/0009-0009-4062-6132}{\includegraphics[scale=0.06]{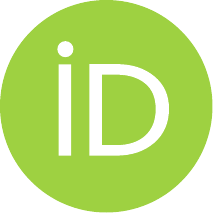}\hspace{1mm}Harry Winston Sullivan} \\ 
	Department of Chemical Engineering and Material Science\\
	University of Minnesota - Twin Cities\\
	Minneapolis, MN \\
	\And
    \href{https://orcid.org/0009-0006-8620-753X}{\includegraphics[scale=0.06]{orcid.pdf}\hspace{1mm}Matej Cervenka} \\
	Institute of Organic Chemistry and Biochemistry\\
    Czech Academy of Sciences\\
	Prague, CZH \\
    \And
    	\href{https://orcid.org/0000-0002-3453-7258}{\includegraphics[scale=0.06]{orcid.pdf}\hspace{1mm}Brennon L. Shanks}\thanks{Corresponding Author: \href{mailto:shanks.brennon@uochb.cas.cz}{shanks.brennon@uochb.cas.cz}} \\
	Institute of Organic Chemistry and Biochemistry\\
    Czech Academy of Sciences\\
	Prague, CZH \\
    \And
	\href{https://orcid.org/0000-0001-9648-6911}{\includegraphics[scale=0.06]{orcid.pdf}\hspace{1mm}Michael P. Hoepfner}\thanks{Corresponding Author: \href{mailto:michael.hoepfner@utah.edu}{michael.hoepfner@utah.edu}} \\
	Department of Chemical Engineering\\
    University of Utah\\
	Salt Lake City, UT \\
}

\begin{document}
\maketitle

\begin{abstract}

We present a nonparametric Bayesian framework to infer radial distribution functions from experimental scattering measurements with uncertainty quantification using non-stationary Gaussian processes. The Gaussian process prior mean and kernel functions are designed to mitigate well-known numerical challenges with the Fourier transform, including discrete measurement binning and detector windowing, while encoding fundamental yet minimal physical knowledge of liquid structure. We demonstrate uncertainty propagation of the Gaussian process posterior to unmeasured quantities of interest. Experimental radial distribution functions of liquid argon and water with uncertainty quantification are provided as both a proof of principle for the method and a benchmark for molecular models. The full implementation is available on GitHub at:\url{https://github.com/hoepfnergroup/LiquidStructureGP-Sullivan}.

\end{abstract}


\newpage
\section{Introduction}

The radial distribution function (RDF), which characterizes the spatial arrangement of atoms, is a cornerstone in liquid state theory that serves as a vital benchmark for molecular simulations. Our understanding of the liquid state relies heavily on established theoretical relationships that link the RDF to thermodynamic properties and interatomic forces. These include the Ornstein-Zernike relation \cite{ornstein_accidental_1914}, Henderson's inverse theorem \cite{henderson_uniqueness_1974}, the Born-Bogilubov-Green-Kirkwood-Yvon hierarchy \cite{hansen_theory_2013}, and Kirkwood-Buff integrals \cite{kirkwood_statistical_1951}, among others. Despite these profound and intricate connections, structure is often relegated to a validation step in molecular modeling, with preference typically given to training force field parameters using macroscopic thermodynamic data \cite{mackerell_jr_empirical_2004} or on interatomic potentials computed from quantum mechanical methods \cite{deringer_gaussian_2021}. While both approaches can yield models that accurately reproduce the thermophysical properties of fluids, they often struggle to fully capture structural features observed in experiments \cite{fheaden_structures_2018, cervenka_cation_2025, fan_charge_2025}. We argue that, to more closely align molecular models with the principles of statistical mechanics, greater emphasis should be placed on experimentally derived RDFs in force field optimization and design.

In X-ray and neutron scattering experiments, the observed quantity is the momentum-space static structure factor, from which RDFs are subsequently inferred \cite{willis_experimental_2017}. For single atom systems, the static structure factor is related to the radial distribution function, $g(r)$, via a Fourier transform with radial symmetry, 
\begin{equation}\label{eq:radFT}
      S(q) - 1 =  4\pi\rho\int_{0}^{\infty} (g(r) -1)\frac{\sin(qr)}{qr}r^2 dr
\end{equation}
\noindent where $q$ is the momentum transfer and $\rho$ is the atomic number density. In mixtures or molecular liquids, the total structure factor, $F(q)$, can be expressed as a combination of site-site partial structure factors, $S_{ij}$, between atoms $i,j$ such that,
\begin{equation} \label{eq:faber}
    F(q) = \sum_{i\geq j} [2 - \delta_{ij}]w_{ij}S_{ij}(q)
\end{equation}
\noindent where $w_{ij}$ is a (possibly q dependent in the case of X-rays) weighting factor depending on the scattering length density and atomic concentration of the $i,j$ pair and $\delta_{ij}$ is the Kronecker delta. This linear system, known as the Faber-Ziman decomposition, is ill-posed when the number of measured total structure factors is fewer than the number of unique site-site partial structure factors, which for a system with $N$ distinct atom types has $N(N+1)/2$ unique $S_{ij}$ terms. To constrain this underdetermined linear system, scattering measurements can be performed on isotopologues (systems differing only by isotopic substitutions), which alter scattering length densities without changing the underlying structure. However, in practice, obtaining a sufficient number of isotopologue measurements is often prohibitively expensive in terms of both experimental time and cost of purified isotopes. As a result, solutions to the Faber-Ziman decomposition have historically been approximated using iterative molecular simulation methods to close the linear system with simulated structure data, such as reverse Monte Carlo (RMC) \cite{mcgreevy_reverse_1988} and empirical potential structure refinement (EPSR) \cite{soper_empirical_1996}. 

Assuming that the partial structure factors are known, they can then be Fourier transformed with eq \eqref{eq:radFT} to obtain real space site-site pair distribution functions, $g_{i,j}(r)$, which quantify the atomic density of type $i$ within a spherical shell around any atom of type $j$. The $g_{i,j}(r)$ describes the relative likelihood of finding a neighboring atom at distance $r$; in liquids, it goes to zero at small $r$ due to atom-atom impenetrability, exhibits oscillations that reflect local structure, and approaches unity at large $r$ where correlations vanish.  

Eqs \eqref{eq:radFT} and \eqref{eq:faber} are conceptually appealing, but their practical implementation faces several challenges. First, the finite size of individual neutron detectors constrains structure factor measurements to discrete momentum transfer values, $\Delta q = q_i - q_{i-1}$, which, according to the Peterson-Middleton sampling theorem \cite{petersen_sampling_1962}, can result in aliasing if the sampling efficiency is $<1$. Second, finite detector coverage windows the measurement to a range between some $q_\text{min}$ and $q_\text{max}$, preventing the evaluation of the full integral specified in eq. \eqref{eq:radFT}. Windowing can introduce truncation artifacts (ripples), reduce the real-space resolution, and, when a smooth windowing correction is applied, can artificially broaden the RDF peaks. Finally, measurement uncertainty of neutron counts and momentum transfer positions (\textit{i.e.} time-of-flight uncertainty) introduce noise that can corrupt the underlying signal \cite{neuefeind_nanoscale_2012,shanks_bayesian_2024}. In practice, a discrete radial Fourier transform must be computed over uncertain observations,
\begin{equation}
    g(r) \approx
    1 + \frac{1}{2\pi^2\rho} \sum_{i=1}^N \frac{1}{2}\bigg(S(q_{i-1}) \frac{\sin( q_{i-1}r)}{ q_{i-1}r} q_{i-1}^2- S(q_{i}) \frac{\sin( q_{i}r)}{ q_{i}r} q_{i}^2 \bigg)\Delta q
\end{equation}
\noindent where the sum is from some non-zero $q_\text{min}$ to some finite $q_\text{max} = N \Delta q$. The key problem is that, depending on the degree of undersampling and the choice of window function, the discrete Fourier transform can systematically distort the predicted RDF relative to the ground truth. These distortions, in turn, increase the uncertainty in the inferred fluid structure. This uncertainty may partially explain why scattering data is not more widely used as an optimization target in force field design.

The most well studied problem in prior literature is addressing the $q_\text{max}$ cutoff using so-called modification functions \cite{proctor_comparison_2023}. The essential idea here is to smoothly transition the structure factor from a data dominated section (as measured by the neutron/X-ray detector) to a model driven section (dictated by prior physical knowledge of the structure factor). Modification functions are designed to force the contribution of the experimental data to 0 near $q_\text{max}$, effectively nullifying any features in the data and strictly relying on the physical model alone. Usually, the data is transitioned into a Poisson point process ideal gas model (i.e. $S_\text{Ideal}(q) = 1$)\cite{torquato_hyperuniform_2018}. Mathematically, this modifies the integral of eq \eqref{eq:radFT} into,
\begin{equation}
    g(r) = 1 + \frac{1}{2\pi^2\rho}\int_{0}^{\infty} (S(q)-1)M(q)\frac{\sin(qr)}{qr}q^2 dq \label{eq:modtransform}
\end{equation} 
\noindent where $M(q)$ is the $q$-dependent modification function. Common choices for the modification function are the first Bessel function \cite{lorch_neutron_1969}, second Bessel function \cite{soper_use_2012,soper_extracting_2011}, cosine cutoff \cite{bellissent-funel_neutron_1992}, and dynamic functions \cite{skinner_benchmark_2013}. However, as pointed out by J.E. Proctor and co-workers \cite{proctor_comparison_2023}, eq \eqref{eq:modtransform} is an approximate Bayesian predictive model where the modification function transitions into a prior $S(q)$ model. To see this, one can rewrite eq $\ref{eq:modtransform}$ in the following way,
\begin{equation}
    = 1 + \frac{1}{2\pi^2\rho}\int_{0}^{\infty} \bigg(\underbrace{(S(q)-1)M(q)}_\text{Data Driven Predictive}+ \underbrace{(S_\text{Ideal}(q)-1)(1-M(q))}_\text{Model Driven Predictive}\bigg)\frac{\sin(qr)}{qr}q^2 dq.\label{eq:approxbayestransform}
\end{equation} 
\noindent where we have split the two contributions of the integrand into "data-driven" and "model-driven" parts regulated by the modification function. Here the $M(q)$ is viewed as a discrete posterior probability mass, meaning that all we have done is expressed the structure factor as a weighted mixture of two outcomes, either data or model, at each $q$ value. An extensive analysis of commonly used modification functions can be found in Ref \cite{proctor_comparison_2023}.

While using prior information to constrain the space of possible RDFs is a valuable idea, the formulation above does not naturally support uncertainty quantification in the RDF predictions within a probabilistic framework. As a result, it does not naturally support uncertainty quantification in the RDF predictions. Prior studies have attempted to estimate uncertainty in scattering data by averaging RDF predictions across multiple experiments and computing standard deviations \cite{soper_radial_2013}, by propagating experimental structure factor errors through the Fourier transform \cite{weitkamp_hydrogen_2000}, or by combining both methods \cite{skinner_benchmark_2013}. The primary limitation of simply comparing different datasets or analysis methods is that such estimates become unreliable if all sources share a common systematic error. Similarly, the Fourier transform error propagation method produces the largest uncertainties at small $r$, precisely where theories of interatomic forces dictate that the RDF must vanish. This results in physically unrealistic uncertainty estimates. A more rigorous approach has been introduced in which Bayesian uncertainty quantification is applied to the interatomic potential using experimentally derived RDFs as observations \cite{shanks_accelerated_2024}. However, this parametric approach relies on a predefined functional form for the potential (for example, a Lennard-Jones or Mie potential), inherently constraining the model and introducing unnecessary bias. What is needed, therefore, is a probabilistic framework that can incorporate known physical features of the RDF, such as short-range exclusion and long-range decay, while remaining flexible enough to avoid biases imposed by assumed potential forms or molecular simulation models.

We propose that a mathematically rigorous version of the RDF posterior satisfying these requirements can be computed through the use of Bayesian inference on the experimental structure factor directly. Specifically, by placing a Gaussian process (GP) prior over the experimental structure factor, multiplying it with an appropriate likelihood function, and finally computing the radial Fourier transform over the resulting $q$-space posterior distribution, we obtain a Bayesian posterior distribution on the RDF. The use of a prior regularizes the infinite set of possible functions that could fit the finite observed dataset, while the likelihood serves as a data fit penalty. The resultant posterior distribution on the RDF represents a direct uncertainty quantification over the real-space structure given the momentum space scattering observations.

GPs have been used extensively in solving ill-posed inverse problems in computational chemistry \cite{deringer_gaussian_2021}, including Fourier analysis of noisy and truncated signals which plagues the scattering problem \cite{ambrogioni_integral_2018}. A GP based approach was recently developed to analyze small-angle neutron scattering data, demonstrating that GP predictions can optimize neutron beamtime usage and increase experimental throughput without compromising data quality \cite{tung_unlocking_2025}. Furthermore, GPs naturally resolve many of the current challenges of scattering analysis cited earlier. For example, they can infer the structure factor on a continuum of momentum values with a domain consistent with the radial Fourier transform (from $q = 0$ to $\lim_{q\rightarrow \infty} S(q)$). As long as the GP mean and kernel selection are physics-informed and flexible enough to represent the data, Bayesian inference will be robust up to available experiments and our theoretical understanding of structure. Such a framework is more elegant and satisfactory than, say, neural networks or other black-box machine learning tools, that often ignore expert knowledge and do not have UQ built-in to their mathematical formalism. The GP framework therefore supports inference while maintaining transparency and expert interpretability. 

In this study, we present a probabilistic machine learning framework to estimate total or partial RDFs with UQ/P from background and inelastic corrected scattering data using non-stationary GP regression. We show how non-stationary GPs with a physics-informed mean and kernel conditioned on experimental scattering data enables the complete reconstruction of the atomic structure from both simulation and experimentally derived total structure factors. The mean and kernel selection reflect simple and indisputable properties of the RDF, including the correct limiting behaviors for realistic bulk fluids ($\lim_{r \rightarrow 0} g(r) = 0$, $\lim_{r \rightarrow \infty} g(r) = 1$), continuity and differentiability, and the presence of non-bonded and bonded contributions.

As test cases for the non-stationary GP model, we performed RDF inference for a simple liquid (argon) and a complex liquid (water) from structure factors derived from both simulation and experiment. All inferred RDF distributions are free from spurious Fourier artifacts and preserve tailing behaviors as dictated by the non-stationary kernel. For liquid argon, we find that the non-stationary GP prediction shows near perfect agreement to a gold-standard neutron scattering analysis from Yarnell \cite{yarnell_structure_1973}. For water structure obtained from a classical water model with flexible bonds, the non-stationary GP regression reconstructs the ground truth data even with significant noise introduced to the structure factor signal. Once the non-stationary GP was validated, we investigated an X-ray scattering dataset of liquid water \cite{skinner_benchmark_2013} to obtain a novel interpretation of the OO RDF distribution that can be compared with molecular models of the liquid structure. 

At first glance, it might seem relatively uninteresting to perform Bayesian inference over the RDF prediction. However, this type of computation can serve as a key result for validation of molecular dynamics simulation \cite{skinner_benchmark_2013,amann-winkel_x-ray_2016} and help unlock emerging methods in computational chemical physics. For instance, access to a RDF distribution as a GP could serve as a link between the increasingly popular Gaussian approximation potential (GAP) framework \cite{deringer_machine_2019} and experimental scattering data. Additionally, force field optimization algorithms such as structure optimized potential refinement (SOPR) \cite{shanks_transferable_2022,shanks_experimental_2025}, which models the interatomic potential as a GP, can now propagate uncertainty directly from experimental observation into the estimation of interatomic potentials. The same is true of parametric Bayesian force field optimization, which can be employed with the methods presented here to estimate how well a given molecular model represents complex experimental data. Finally, as Bayesian interpretations become more frequently integrated into chemical physics, these approaches will be necessary to understand model uncertainties with respect to experiments, a critical step of the scientific method that has been largely under-reported in the existing literature due to a lack of rigorous approaches to estimate uncertainty in complex experimental observables. This framework therefore allows us to leverage all available data without throwing away established physical knowledge accumulated through generations of scientific discovery. 

\section{Theory and Methods}

\begin{figure}
    \centering
    \includegraphics[width=0.9\linewidth]{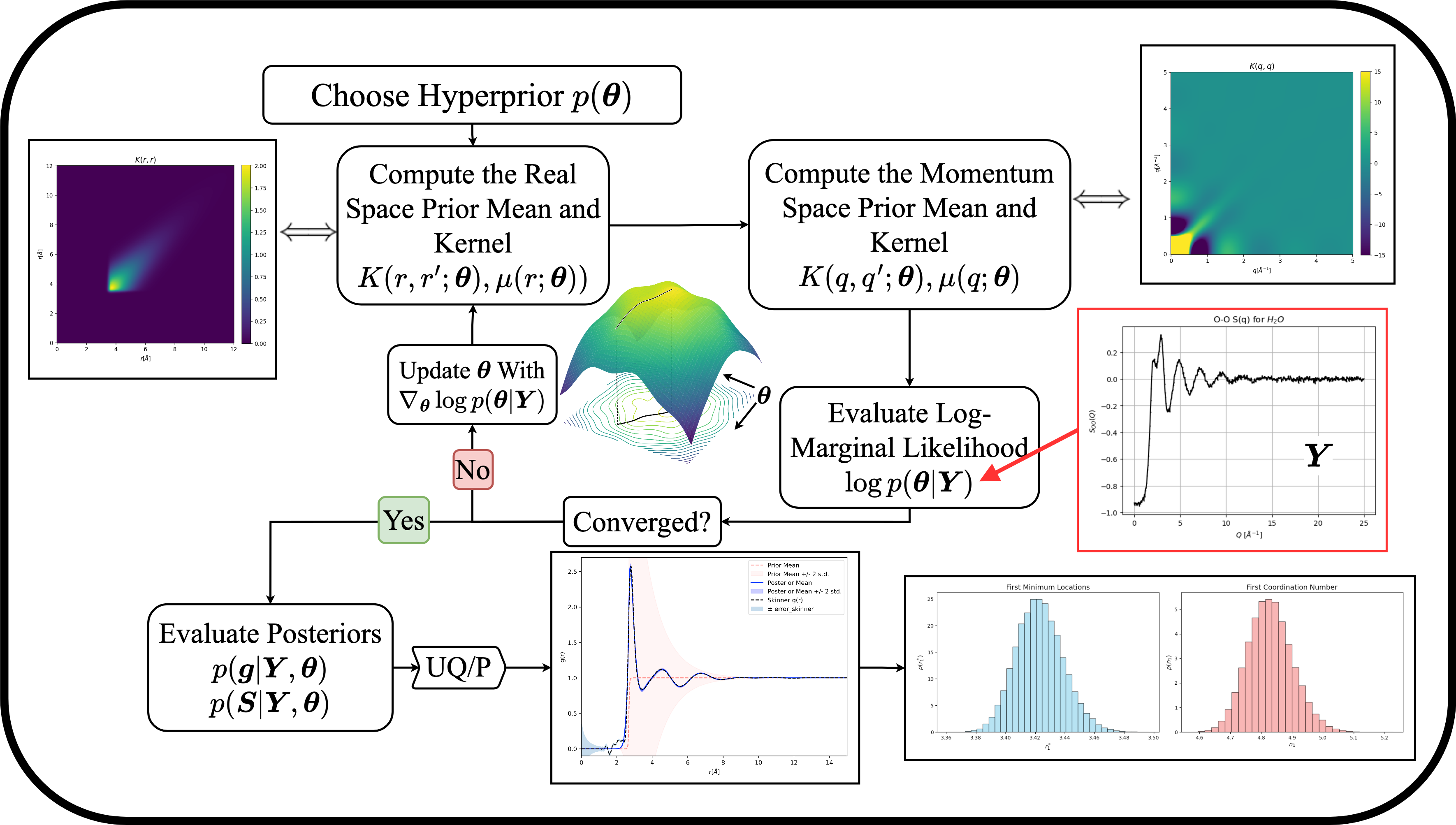}
    \caption{A flowchart corresponding to the GPFT algorithm applied to scattering data.}
    \label{fig:ALG}
\end{figure}

Consider a structure factor, which is unknown to us, that lives within a \textit{distribution} of possible functions that are known to obey specific physical characteristics. We can write this mathematically in the context of a GP by stating that any evaluation (or set of evaluations) of the unknown function, $S(q)$, is distributed as a multivariate Gaussian,
\begin{equation}
    S(q) \sim \mathcal{GP}(\mu(q), K(q,q'))
\end{equation}
\noindent where the mean, $\mu(x)$, represents the \textit{a priori} expected value of the function at each point in the input space, and the covariance (or kernel) function, $K(q,q')$, represents the relatedness of the output quantities with respect to the process inputs. The mean and kernel constrain the set of possible functions that could represent the experimental observation to those consistent with physical intuition. The non-stationary behavior of such a GP refers to the fact that the Fourier transform of the true structure factor has limiting behavior with certainty (i.e., $\lim_{r \rightarrow 0} g(r) = 0$, $\lim_{r \rightarrow \infty} g(r) = 1$), meaning that the functional distribution has covariances that change with respect to its inputs \cite{heinonen_non-stationary_2016}. 

A physics-informed GP model for the structure factor enables Bayesian inference of a structure factor posterior distribution conditioned on experimental scattering data. This posterior reflects uncertainties from both known physical principles and experimental observations, offering a reliable and robust estimation of the uncertainty in the liquid structure. Finally, due to the linearity of the radial Fourier transform, this uncertainty can be propagated into the prediction of the RDF and subsequently compared to molecular simulation predictions. This rigorous representation of our current knowledge of the liquid structure serves as a powerful validation tool for molecular models and helps pinpoint key measurements necessary to resolve gaps in our understanding of the organization of molecules in liquids.

\subsection{The Gaussian Process Model}

Bayes' theorem provides a natural route to compute the structure factor \textit{posterior}, $p(\boldsymbol{S}|\boldsymbol{Y},\boldsymbol{\theta})$, according to Bayes theorem,
\begin{equation}
    p(\boldsymbol{S}|\boldsymbol{Y},\boldsymbol{\theta}) = \frac{p(\boldsymbol{Y}|\boldsymbol{S},\boldsymbol{\theta})p(\boldsymbol{S}|\boldsymbol{\theta})}{p(\boldsymbol{Y}|\boldsymbol{\theta})} =  p(\boldsymbol{Y}|\boldsymbol{S},\boldsymbol{\theta})p(\boldsymbol{S}|\boldsymbol{\theta})
    \bigg(\int p(\boldsymbol{Y}|\boldsymbol{S},\boldsymbol{\theta})p(\boldsymbol{S}|\boldsymbol{\theta}) d\boldsymbol{S}\bigg)^{-1} \label{eq:bayes}
\end{equation}
\noindent where  $p(\boldsymbol{S}|\boldsymbol{\theta})$, $p(\boldsymbol{Y}|\boldsymbol{S},\boldsymbol{\theta})$, and $p(\boldsymbol{Y}|\boldsymbol{\theta})$ are the prior, likelihood, and model evidence, respectively. Here $\boldsymbol{Y}$ is the set of experimental observations, $\boldsymbol{S}$ is the value of the structure factor due to some inducing index vector, and $\boldsymbol{\theta}$ is the set of GP hyperparameters.  

The \textit{GP prior} over the inducing index vector of momentum transfer values $\boldsymbol{q}$ is defined by a mean, $\mu(\boldsymbol{q}) = \boldsymbol{\mu}_{\boldsymbol{q}}$, and kernel, $K(\boldsymbol{q},\boldsymbol{q}) = \hat{\boldsymbol{K}}_{\boldsymbol{q},\boldsymbol{q}}$, function which when evaluated on the index set produce a vector and a matrix respectively,
\begin{align}
    p(\boldsymbol{S}|\boldsymbol{\theta}) = (2\pi)^{-d/2}\det|\hat{\boldsymbol{K}}_{\boldsymbol{q},\boldsymbol{q}}|^{-\frac{1}{2}}\exp((\boldsymbol{S} - \boldsymbol{\mu}_{\boldsymbol{q}})^T \hat{\boldsymbol{K}}_{\boldsymbol{q},\boldsymbol{q}}^{-1}(\boldsymbol{S} - \boldsymbol{\mu}_{\boldsymbol{q}}))
\end{align}
\noindent where the determinant is required due to the non-diagonal covariance between the latent function values. Note that this quantity is just a GP representation of the structure factor distribution before it has seen any data.

The \textit{likelihood} is the probability that the observed data is generated by a particular instance of $\boldsymbol{S}$. Assuming that the structure factor has approximately spatially uncorrelated and constant Gaussian noise (which is the case for reactor source neutron scattering \cite{willis_experimental_2017}), an appropriate likelihood is a homoscedastic normal distribution,
\begin{align}
    p(\boldsymbol{Y}|\boldsymbol{S},\boldsymbol{\theta}) = (2\pi\omega^2)^{-d/2}\exp((\boldsymbol{S} - \boldsymbol{Y})^T (\omega^2\hat{\boldsymbol{I}})^{-1}(\boldsymbol{S} - \boldsymbol{Y})) \label{eq:LH}
\end{align}
\noindent where $\omega$ is a noise parameter, $d$ is the number of observed data points and $\hat{\boldsymbol{I}}$ is the identity matrix. Modified versions of this expression would be required for systems with significant heteroscedastic noise, such as spallation source neutron instruments. In this case, the likelihood can be generalized by replacing the scalar variance $\omega^2$ with a function $\omega(\boldsymbol{q})^2$ varying along the diagonal of the covariance matrix while remaining jointly Gaussian. In general, as long as the observations are independent the likelihood can represent any noise distribution, underscoring the flexibility of the Bayesian framework \cite{li_improving_2023}. Note that the dependence of these expressions on the hyperparameters $\boldsymbol{\theta}$ stems from the underlying kernel and mean functions used to evaluate $\boldsymbol{S}$.

Finally, the conjugacy of Gaussian distributions for both the prior and likelihood enables analytical integration of the \textit{model evidence} (also known as the marginal likelihood) \cite{bishop_pattern_2006,rasmussen_gaussian_2006}, and, upon taking the log to improve numerical stability, gives,
\begin{align}
    \log p(\boldsymbol{Y}|\boldsymbol{\theta}) = -\frac{1}{2}(\boldsymbol{Y}-\boldsymbol{\mu}_{\boldsymbol{q}})^T(\hat{\boldsymbol{K}}_{\boldsymbol{q},\boldsymbol{q}} + \omega^2\hat{\boldsymbol{I}})^{-1}(\boldsymbol{Y}-\boldsymbol{\mu}_{\boldsymbol{q}}) - \frac{1}{2}\log\det|\hat{\boldsymbol{K}}_{\boldsymbol{q},\boldsymbol{q}} +\omega^2\hat{\boldsymbol{I}} | - \frac{d}{2}\log{2\pi}. \label{eq:LMLH}
\end{align}
Combining these expressions into eq \eqref{eq:bayes}, the posterior distribution over the latent function $\boldsymbol{S}$ evaluated at some $m$ sized index vector $\boldsymbol{q}^*$ is then,
\begin{align}
     & p(\boldsymbol{S}|\boldsymbol{Y},\boldsymbol{\theta}) = (2\pi)^{-m/2}\det|\hat{\boldsymbol{\Sigma}}_{\text{Post}}|^{-\frac{1}{2}}\exp((\boldsymbol{S} - \boldsymbol{\mu}_{\text{Post}})^T \hat{\boldsymbol{\Sigma}}_{\text{Post}}^{-1}(\boldsymbol{S} - \boldsymbol{\mu}_{\text{Post}})) \label{eq:post_sq}
\end{align}
\noindent where the posterior mean and variance are given by,
\begin{align}
     & \boldsymbol{\mu}_{\text{Post}} = \boldsymbol{\mu}_{\boldsymbol{q}^*} + \hat{\boldsymbol{K}}_{\boldsymbol{q}^*,\boldsymbol{q}} (\hat{\boldsymbol{K}}_{\boldsymbol{q},\boldsymbol{q}} + \omega^2\hat{\boldsymbol{I}})^{-1} (\boldsymbol{Y} - \boldsymbol{\mu}_{\boldsymbol{q}}) \label{eq:SQ_post_mu} \\
     & \hat{\boldsymbol{\Sigma}}_{\text{Post}} = \hat{\boldsymbol{K}}_{\boldsymbol{q}^*,\boldsymbol{q}^*} - \hat{\boldsymbol{K}}_{\boldsymbol{q}^*,\boldsymbol{q}} (\hat{\boldsymbol{K}}_{\boldsymbol{q},\boldsymbol{q}} + \omega^2\hat{\boldsymbol{I}})^{-1} \hat{\boldsymbol{K}}_{\boldsymbol{q},\boldsymbol{q}^*} \label{eq:SQ_post_sigma}
\end{align}

\subsection{Posterior Estimation of the Radial Distribution Function}\label{sec:post_estimate}

The next step is to propagate uncertainty from the structure factor distribution into real-space. By the fluctuation-dissipation theorem, the RDF is related to the structure factor through a 3D Fourier transform, which, assuming spherical symmetry, can be written as the well-known radial Fourier transform (rFT) which we denote $\Tilde{\boldsymbol{\mathcal{H}}}$,
\begin{align}
    \Tilde{\boldsymbol{\mathcal{H}}}_q[f(q)] = \frac{1}{2\pi^2\rho}\int_0^\infty f(q) \frac{\sin(qr)}{qr} q^2 dq, \quad \Tilde{\boldsymbol{\mathcal{H}}}^{-1}_r[f(r)] = 4\pi\rho\int_0^\infty f(r) \frac{\sin(qr)}{qr} r^2 dr.
\end{align}
\noindent which maps a function of $r$ to a function of $q$. Notably, the inverse of the rFT proportional to the operator is itself up to a proportionality constant ($\Tilde{\boldsymbol{\mathcal{H}}} =  8\pi^3\rho^2\Tilde{\boldsymbol{\mathcal{H}}}^{-1}$), see Supporting Information Section C for details) and is linear with respect to the input function. This operator is related to the Hankel transform \cite{baddour_theory_2015}.
The RDF structure factor relationship is then,
\begin{align}
     S(q) = 1+\Tilde{\boldsymbol{\mathcal{H}}}^{-1}_r[g(r)-1],\quad g(r) = 1 + \Tilde{\boldsymbol{\mathcal{H}}}_q[S(q)-1].\label{eq:transform}
\end{align}
At first glance, it may seem unclear how to apply eq \eqref{eq:transform} to a distribution of structure factors; however, the normality of the GP, in tandem with the linearity of the operator, can alleviate nearly all of the difficulty since the resulting distribution is trivially Gaussian. This nice property is due to the well-known fact that the linear transformation of a finite dimensional Gaussian distribution is again Gaussian,
\begin{align}
    & \boldsymbol{z}  \sim \mathcal{N}(\boldsymbol{\mu},\hat{\boldsymbol{\Sigma}}) \\
    & \implies \hat{\boldsymbol{A}}\boldsymbol{z}  \sim \mathcal{N}(\hat{\boldsymbol{A}}\boldsymbol{\mu},\hat{\boldsymbol{A}}\hat{\boldsymbol{\Sigma}}\hat{\boldsymbol{A}}^T)\label{eq:linear_transformation}
\end{align}
\noindent where $\hat{\boldsymbol{A}}$ is a linear operator acting on a finite dimensional vector $\boldsymbol{z}$. Assuming that the linear operator is bounded and densely defined, the same property holds for GPs \cite{matsumoto_images_2024}. This approach is often leveraged in the analysis of partial differential equations \cite{swiler_survey_2020} and can be applied to the rFT integral operator defined in eq \eqref{eq:transform}.

Eq \eqref{eq:linear_transformation} has important implications for relating kernels between the Fourier duals of momentum- and real-space. For example, we can now construct new kernels in the Fourier dual space by applying the linear rFT operator,
\begin{align}
    & K(r,r') = \text{cov}(g(r),g(r') = \Tilde{\boldsymbol{\mathcal{H}}}_q[\Tilde{\boldsymbol{\mathcal{H}}}_{q'}[K(q,q')]]  \label{eq:Krr} \\ 
    & K(r,q') =  \text{cov}(g(r),S(q')) =\Tilde{\boldsymbol{\mathcal{H}}}_q[K(q,q') ]\label{eq:Krq}\\ 
    & K(q,r') = \text{cov}(S(q),g(r')) = \Tilde{\boldsymbol{\mathcal{H}}}^{-1}_{r}[K(r,r')]\label{eq:Kqr} \\ 
    & K(q,q') = \text{cov}(S(q),S(q')) =\Tilde{\boldsymbol{\mathcal{H}}}^{-1}_r[\Tilde{\boldsymbol{\mathcal{H}}}^{-1}_{r'}[K(r,r')]]\label{eq:Kqq} %
\end{align}
In essence, the RDF posterior distribution reflects correlations between observed data in $q$- space projected into $r$-space, giving the overall probability of the RDF $\boldsymbol{g}$ evaluated on a $n$ sized index vector $\boldsymbol{r}$ as,
\begin{align}
     & p(\boldsymbol{g}|\boldsymbol{Y},\boldsymbol{\theta}) = (2\pi)^{n/2}\det|\hat{\boldsymbol{\Sigma}}_{\text{Post,RDF}}|^{-\frac{1}{2}}\exp((\boldsymbol{g} - \boldsymbol{\mu}_{\text{Post,RDF}})^T \hat{\boldsymbol{\Sigma}}_{\text{Post}}^{-1}(\boldsymbol{g} - \boldsymbol{\mu}_{\text{Post,RDF}})) \label{eq:post_rdf}
\end{align}
\noindent with posterior mean and variance,
\begin{align}
     & \boldsymbol{\mu}_{\text{Post,RDF}}  = \boldsymbol{\mu}_{\boldsymbol{r}} + \hat{\boldsymbol{K}}_{\boldsymbol{r},\boldsymbol{q}} (\hat{\boldsymbol{K}}_{\boldsymbol{q},\boldsymbol{q}} + \omega^2\hat{\boldsymbol{I}})^{-1} (\boldsymbol{Y} - \boldsymbol{\mu}_{\boldsymbol{q}}) \label{eq:RDF_post_mu}\\
     & \hat{\boldsymbol{\Sigma}}_{\text{Post,RDF}} = \hat{\boldsymbol{K}}_{\boldsymbol{r},\boldsymbol{r}} - \hat{\boldsymbol{K}}_{\boldsymbol{r},\boldsymbol{q}} (\hat{\boldsymbol{K}}_{\boldsymbol{q},\boldsymbol{q}} + \omega^2\hat{\boldsymbol{I}})^{-1} \hat{\boldsymbol{K}}_{\boldsymbol{q},\boldsymbol{r}} \label{eq:RDF_post_sigma}
\end{align}
\noindent where $\boldsymbol{\mu}_{\boldsymbol{r}}$ is just $\Tilde{\boldsymbol{\mathcal{H}}}_q[\mu(q) - 1]$ evaluated at $\boldsymbol{r}$. Formally, these expressions may also be obtained by through application of the rFT $\Tilde{\boldsymbol{\mathcal{H}}}_q$ operator to the $S(q)$ posterior evaluated at a single inducing point $q$. 

While the above method works in theory, not all of the integrals are analytically tractable and conducive to pen and paper computation. Indeed, it is more practical to use an approximate operator, $\Tilde{\boldsymbol{\mathbb{H}}}$, computed with simple numerical quadrature, 
\begin{align}
    \Tilde{\boldsymbol{\mathcal{H}}}_q [f(q)]\approx \sum_{i=1}^N \frac{1}{4\pi^2\rho}\bigg(f(q_{i-1}) \frac{\sin( q_{i-1}r)}{ q_{i-1}r} q_{i-1}^2- f(q_{i}) \frac{\sin( q_{i}r)}{ q_{i}r} q_{i}^2 \bigg)\Delta q = \Tilde{\boldsymbol{\mathbb{H}}}_q [f(\boldsymbol{q})] \label{eq:drft}
\end{align}
\noindent where the grid of $q$ values is over the range of the integral. Note that the approximate operator $\Tilde{\mathbb{H}}$ acts on a discretized grid of function values $f(\boldsymbol{q})$ and produces a single number which corresponds to the implicit radial argument $r$. The choice of grid spacing can affect the resulting RDFs and uncertainty predictions, therefore care must be taken to ensure the grid resolves all relevant features of the kernel and mean functions used. Details on the exact grid spacing and limits used are provided in the Appendix. This approximate rFT operator retains linearity while generalizing to custom GP prior means and kernels that do not have analytical radial Fourier transforms. This strategy holds connections with typical Bayesian quadrature techniques \cite{ohagan_bayeshermite_1991}. The inverse is discretized similarly with an alternate prefactor.

\subsection{Designing a Gaussian Process for Liquid Structure Factors}\label{sec:kernel}

The crux of designing any GP model is choosing an appropriate prior, which for a GP is fully specified by its mean and kernel functions and their corresponding hyperparameters. This step is also the most critical for enforcing physical behaviors and constraints in the GP regression. 

Because physical correlations are less transparent in momentum space, it is more intuitive to impose constraints directly on the real-space RDF, where structural features are more easily interpreted and well-understood. Given a real-space mean $\mu(r)$ and kernel $K(r,r')$, we can then perform a radial Fourier transform using one of the techniques from the previous section to obtain $K(q,q)$ as well as the log marginal likelihood in eq \eqref{eq:LMLH}. Past work has shown that capturing the limiting behaviors correctly can greatly improve the transform procedure \cite{proctor_comparison_2023}. Therefore, it is crucial to ensure proper boundary behaviors in the GP prior. We know there must be an excluded volume, as well as a trend towards 1 at the limit, to preserve the overall density of the fluid. Mathematically, these boundary conditions are expressed as,
\begin{align}
    \lim_{r\rightarrow 0} g(r) = 0 \tab \lim_{r\rightarrow \infty} g(r) = 1
\end{align}
\noindent which can be incorporated into the GP model directly using a non-stationary kernel. 

\subsubsection{Non-Stationary Kernel Selection}

Unlike stationary kernels, which assume that the covariance only depends on the distance between inducing input locations ($k(r,r') = k(|r - r'|)$), non-stationary kernels allow the covariance to change across different regions of the input space. A cleverly designed non-stationary kernel can improve the model’s predictive power by ensuring that the GP adheres to known physical constraints. To see why this is the case, consider that when the kernel evaluation approaches zero, the GP distribution tends towards the mean (\textit{c.f.} eqs \eqref{eq:SQ_post_mu} and \eqref{eq:SQ_post_sigma}). This provides a natural strategy for enforcing the boundary conditions: force the mean to a known limiting behavior while forcing the covariance to vanish. Specifically, the limits we are after are,
\begin{align}
    & \lim_{r \text{ or } r' \rightarrow 0 \text{ or } \infty} K(r,r') = 0 \label{eq:kernel_limit}, \tab \lim_{r \rightarrow 0} \mu(r) = 0, \tab \lim_{r \rightarrow \infty} \mu(r) = 1
\end{align}
\noindent ensuring that the kernel captures localized variations away from the boundaries ($r = 0$ and $r \to \infty$), while the mean function encodes the global boundary behaviors. 

Although the RDF is technically a map from $\mathbb{R}^+$ to $\mathbb{R}^+$, the GP itself is not restricted to this domain. To account for this, we impose symmetry with respect to $r=0$ by symmetrizing the kernel,
\begin{align}
    K_\text{Sym.}(r,r') = K(r,r') + K(-r,r').
\end{align}
Symmetrizing the kernel prevents artificial asymmetries in the model and ensures that the process behaves consistently across the full input space. Additionally, we know that $g(r)$ must be continuous and differentiable, as it is required to belong to the radially symmetric Schwartz space to be Fourier transformable \footnote{Notably one can extend the domain of the FT to the tempered distributions, we do not consider such cases in this work.}. Therefore, we based our kernel on the widely used squared exponential kernel, but with non-stationary behavior introduced through $r$-dependent length scale $\ell(r)$ and a width scale $\sigma(r)$ functions. The kernel of this type is known as the Gibbs kernel \cite{gibbs_bayesian_1997}, which is highly flexible and allows for spatially varying properties,
\begin{align}
    K(r,r') = \sigma(r)\sigma(r')\sqrt{\frac{2\ell(r)\ell(r')}{\ell^2(r) + \ell^2(r')}}\exp\bigg(\frac{-(r-r')^2}{\ell^2(r) + \ell^2(r')}\bigg).
\end{align}
The flexibility of the Gibbs kernel makes it particularly well-suited for systems where the properties of the process change over space in a known way. By parameterizing $\sigma(r)$ and $\ell(r)$ using a chosen functional form, we can further incorporate known behaviors and enhance generalizability. Following the strategy outlined above, we aim to embed as much physically relevant behavior as possible into $\mu(r)$, while selecting $\sigma(r)$ and $\ell(r)$ to account for deviations from the mean.

In the fluid structures of concern to this work, it is atypical see large length scale changes as a function of $r$ (with the exception of the bonded vs non-bonded structure handled in the mean). This allows us to choose $\ell(r)$ to be a constant. The limiting behavior of the kernel at large or small inputs is then encoded within $\sigma(r)$. By ensuring that $\sigma(r)$ tends to zero as $r$ tends to $0$ or $\infty$ the kernel will satisfy eq \eqref{eq:kernel_limit}. Simple functional forms that satisfy these constraints are a constant length scale function and a decaying sigmoid for the width function,
\begin{align}
    \ell(r) = \ell, \tab \sigma(r) =\frac{\text{Max}\cdot\exp(\text{Decay}\cdot\text{Loc})}{1+\exp(-\text{Slope}\cdot(r-\text{Loc}))}\exp(-r\cdot\text{Decay}))\label{eq:width}
\end{align}
\noindent where the hyperparameters $\text{Max}$, $\text{Decay}$, $\text{Loc}$, and $\text{Slope}$ control the height, decay rate, peak location, and sharpness of the peak in the sigmoid respectively. 

The presented phenomenological kernel represents an Occam's razor strategy to kernel design. However, while this kernel satisfies the relevant physical constraints, alternative functional forms with comparable properties and varying levels of rigor are certainly possible. In principle, both $\ell(r)$ and $\sigma(r)$  could be derived from first principles, modeled as latent functions (e.g., GPs with their own mean and kernel structures), or selected phenomenologically, as we do here. Conveniently, the Bayesian GP framework provides a principled way to compare kernels through the computation of Bayes factors, which are ratios of their respective marginal likelihoods, $BF = \frac{p(y|K_1)}{p(y|K_2)}$. The higher the value of the Bayes factor, the more the data supports $K_1$. To compute the marginal likelihood for a candidate kernel, hierarchical inference over the kernel hyperparameters must be performed to fully account for uncertainty. Although this approach is rigorous, its high computational cost places it beyond the scope of the present study. Nonetheless, it remains a promising direction for future work in physics-informed kernel design and selection.

\subsubsection{Mean Selection}

The simplest information to include in the mean $\mu$ is the hard particle-like repulsive shell and bond information. In simulations, bonds are often modeled using a harmonic oscillator, resulting in a sharp, an approximately Gaussian peak in $g(r)$. This leads to the bonded portion of the mean being represented as a sum of Normal distributions,
\begin{align}\label{mean_term}
    \mu_{\text{Bonded}}(r) = \sum_{b=1}^{B} h_b\mathcal{N}(r|r_b,s_b).
\end{align}
Here, the sum is taken over each unique structural peak in the particular molecule. For instance, when studying the oxygen-hydrogen correlation of water, we would expect at least one peak corresponding to the oxygen-hydrogen bond. However, for larger molecules, the situation can become more complex. Consider the hydrogen-hydrogen correlations in benzene. Although each hydrogen atom is exclusively bonded to carbon, we still observe bond-like peaks in $g(r)$ due to the intramolecular hydrogen atoms still being in proximity with one another. These manifest as normal peaks as if they were directly bonded. In principle one could then relate the parameter $r_b$ to the equilibrium bond lengths, $s_b$ to the strength of the bonds, and $h_b$ to the typical number of atoms at the distance $r_b$. The excluded volume part is then represented as a simple sigmoid. This choice aligns with the limit behavior of the mean outlined above,
\begin{align}
     \mu_{\text{Non-Bonded}}(r) = \frac{1}{1+\exp(-s_0(r-r_0))}.
\end{align}
Overall, the mean function for the GP model is then,
\begin{align}
    \mu(r) = \mu_{\text{Bonded}}(r) + \mu_{\text{Non-Bonded}}(r) \\
    = \sum_{b=1}^{B} h_b\mathcal{N}(r|r_b,s_b) + \frac{1}{1+\exp(-s_0(r-r_0))}
\end{align}
The importance of the bonded term in the structure factor is critical for capturing the long-range features in momentum-space. Indeed, most of the long-range oscillatory behavior observed in momentum-space arises from this component of the mean, which becomes apparent when considering the rFT of a shifted Gaussian,
\begin{align}
    \boldsymbol{\mathcal{H}}_r^{-1}\bigg[\frac{1}{\sqrt{2\pi\sigma^2}}\exp\bigg(-\frac{(r-r_0)^2)}{2\sigma}\bigg)\bigg] = 4\pi\rho \int_0^\infty \frac{\sin(qr)}{qr\sqrt{2\pi\sigma^2}} \exp\bigg(-\frac{(r-r_0)^2)}{2\sigma}\bigg) r^2dr.
\end{align}
Assuming the mean is sufficiently far from the origin and the distribution is narrow enough to contribute negligibly for $r<0$ (which is appropriate for chemical bonds), we can extend the integration bounds to the entire real line. This transformation recasts the integral as an expectation value,
\begin{align}
   \boldsymbol{\mathcal{H}}_r^{-1}\bigg[\frac{1}{\sqrt{2\pi\sigma^2}}\exp\bigg(-\frac{(r-r_0)^2)}{2\sigma}\bigg)\bigg] \approx4\pi\rho \frac{\expec_{r\sim\mathcal{N}(r_0,\sigma^2)}\bigg[\sin(qr)qr \bigg]}{q^2}.
\end{align}
Next, by introducing the change of variable $y = qr$, applying Euler's identity $\operatorname{Im}[e^{iy}]=\sin(y)$, completing the square in the exponent, and identifying a new Gaussian with a complex mean, the resulting expectation yields an approximate, yet analytical, expression for the bonded portion of the structure factor mean,
\begin{align}
    \mu_{Bonded} = \frac{4\pi\rho\exp\left({-\frac{(q\sigma)^2}{2}}\right)\left( (q\sigma)^2 \cos(qr_0) + q r_0 \sin(qr_0) \right)}{q^2}.
\end{align}
A closer look reveals that the dominant term decays as $\exp\left({-(q\sigma)^2}\right)/q$, with the standard deviation $\sigma$ controlling the decay rate. A larger $\sigma$ implies a looser, more flexible bond with the central atom, suggesting that the high-$q$ decay rate reflects the bond’s strength.

\subsubsection{Hyperparameter Optimization}

Machine learning inevitably comes down to inferring a robust set of hyperparameters conditioned on available observations, and non-stationary GP regression is no exception. In a fully Bayesian formalism, one can infer hyperparameter posterior distributions and propagate their uncertainty to the GP model via hierarchical Bayesian inference. The first step is a Bayesian inversion of the model evidence, $p(\boldsymbol{Y}|\boldsymbol{\theta})$, to give the conditional probability of the hyperparameters given available data, $p(\boldsymbol{\theta}|\boldsymbol{Y})$. Conceptually, we are asking \textit{what is the probability that the hyperparameters in our prior model generated the training data?} Using Bayes' theorem, we can write the hyperposterior probability in log space as,
\begin{align}
     \log p(\boldsymbol{\theta}|\boldsymbol{Y}) = -\log(Z)  -\underbrace{\frac{1}{2}(\boldsymbol{Y}-\boldsymbol{\mu}_{\boldsymbol{q}})^T(\hat{\boldsymbol{K}}_{\boldsymbol{q},\boldsymbol{q}} + \omega^2 \hat{\boldsymbol{I}})^{-1}(\boldsymbol{Y}-\boldsymbol{\mu}_{\boldsymbol{q}})}_{\textit{Data Fit}} -\underbrace{ \frac{1}{2}\log\det|\hat{\boldsymbol{K}}_{\boldsymbol{q},\boldsymbol{q}}|}_{\textit{Complexity Penalty}} - \frac{d}{2}\log{2\pi} +\underbrace{\log p(\boldsymbol{\theta})}_{\textit{Prior Belief}}\label{eq:hp_post}
\end{align}
\noindent where $Z$ is a normalization constant. By inspection, it is clear that this expression represents a compromise between a data fit and a complexity penalty. As the model improves at matching the training data the \textit{Data Fit} term decreases, overall increasing the log probability. Simultaneously, as the kernel becomes more complex, the determinant in the \textit{Complexity Penalty} will increase, in turn decreasing the log probability. The determinant increase is associated with a larger volume spanned by the eigenvectors of the kernel matrix. The overall expression becomes a competition between these two terms, resulting in an Occam's razor effect that pushes the model towards the simplest explanation of the dataset.

Unfortunately, exact computation of the log probability is not tractable for generic choices of $p(\boldsymbol{\theta})$, $\hat{\boldsymbol{K}}_{\boldsymbol{q},\boldsymbol{q}}$, and $\boldsymbol{\mu}_{\boldsymbol{q}}$. The root of the problem lies in the computation of the normalization constant $Z$ via integration of $p(\boldsymbol{Y}|\boldsymbol{\theta})p(\boldsymbol{\theta})$ over $\boldsymbol{\theta}$, which becomes computationally infeasible even for relatively simple GPs. Therefore, a point estimate of $\boldsymbol{\theta}$ was inferred and used for subsequent GP modeling by maximizing the log hyperposterior with gradient ascent applied to eq \eqref{eq:hp_post}. This process is outlined in the primary loop shown in figure \ref{fig:ALG}.

Utilization of one of the transformation techniques from Section \ref{sec:post_estimate} allows us to rewrite the optimization target stemming from eq \eqref{eq:hp_post} as,
\begin{align}
    \argmax_{\boldsymbol{\theta}}\bigg[ &\frac{1}{2}\left(\boldsymbol{Y}-\Tilde{\boldsymbol{\mathbb{H}}}^{-1}_r[\boldsymbol{\mu}_{\boldsymbol{r}}]\right)^T
    \left(\Tilde{\boldsymbol{\mathbb{H}}}^{-1}_r\left[\Tilde{\boldsymbol{\mathbb{H}}}^{-1}_r [\hat{\boldsymbol{K}}_{\boldsymbol{r},\boldsymbol{r}}]\right] + \omega^2 \hat{\boldsymbol{I}}\right)^{-1}
    \left(\boldsymbol{Y}-\Tilde{\boldsymbol{\mathbb{H}}}_r^{-1}[\boldsymbol{\mu}_{\boldsymbol{r}}]\right) \nonumber \\
    &- \frac{1}{2}\log\det\left|\left(\Tilde{\boldsymbol{\mathbb{H}}}_r^{-1}\left[\Tilde{\boldsymbol{\mathbb{H}}}_r^{-1}[\hat{\boldsymbol{K}}_{\boldsymbol{r},\boldsymbol{r}}]\right] +\omega^2 \hat{\boldsymbol{I}}\right)^{-1}\right|
    +\log p(\boldsymbol{\theta})\bigg]. \label{eq:opt_target}
\end{align}
where the hyperparameter vector is,
\begin{align}
    \boldsymbol{\theta} = [r_0,s_0,h_1,r_1,s_1,h_2,r_2,s_2,\dots h_B,r_B,s_B,\ell,\text{Max},\text{Slope},\text{Loc},\text{Decay},\omega]^T
\end{align}
\noindent The inclusion of $\omega$ is meant to treat the standard deviation in the likelihood as a learnable hyperparameter.

By employing automatic differentiation (AD), we can compute the gradients of the optimization target with respect to the hyperparameters in a precise and computationally efficient manner. This capability allows us to leverage gradient-based optimization algorithms, such as stochastic gradient descent or Adam \cite{bishop_pattern_2006}, to update the hyperparameters iteratively. The primary advantage of using AD is its ability to provide accurate gradients without the need for numerical differentiation, which can be sensitive to perturbations and often requires additional function evaluations. With AD, gradient computation is integrated into the optimization routine, enabling more rapid convergence to optimal hyperparameter values. Furthermore, AD facilitates the inclusion of our custom kernel, enhancing the flexibility of our modeling approach.

One challenge with hyperparameter optimization using this method is the risk of getting stuck in local minima, making a good initial guess crucial to training an accurate model. In machine learning models like neural networks with many, non-interpretable hyperparameters, this is difficult task. However, since the GP mean and kernel functions are based on physical properties, chemical intuition makes finding an initial estimate straightforward. For example, a RDF computed from a single molecular dynamics simulation of the target system can be used to provide a good first guess for the kernel hyperparameters that can be subsequently refined according to the experimental scattering data. This physical interpretability greatly simplifies the hyperparameter training process in the non-stationary GP framework. 

The computational cost of the non-stationary GP method is dominated by this hyperparameter optimization step, which entails repeated cubic-scaling GP evaluations in the number of experimental $q$-space points. The difficulty of optimization grows with the number of hyperparameters, as this typically goes hand-in-hand with a rougher objective function. The numerical Fourier transform unique to this work scales linearly with the number of $r$-space grid points and is negligible in comparison to the matrix inverse. 

Despite the theoretical complexity, in our experiments, we found the optimization to run end-to-end within a couple hours on a laptop given modest-sized datasets (200 to 500 observations). However, for a given set of hyperparameters a single GP inference is of relatively trivial computational cost given standard memory and processor capabilities of modern personal computers up to approximately 10$^3$ observations. The establishment of standardized ranges or additional physical constraints on the hyperparameters would support rapid characterization of numerous samples with the added benefit of uncertainty quantification. Our implementation of the optimization and inference procedure is available on GitHub at \url{https://github.com/hoepfnergroup/LiquidStructureGP-Sullivan}.

\subsection{Coordination Number Analysis}

In addition to enforcing physical constraints during uncertainty quantification, the non-stationary GP framework provides a principled foundation for calculating physical properties and their uncertainty bounds derived from the RDF, such as the coordination number. 

The coordination number is the average number of type~$\beta$ neighbors within a distance~$R$ of a reference atom of type~$\alpha$, given by
\begin{align}
     n(R) = 4\pi \rho_\beta \int_0^R g_{\alpha,\beta}(r) \, r^2 \, dr. \label{eq:coord0}
\end{align}
This standard expression depends sensitively on the choice of integration bounds, which introduces two well-known limitations in conventional scattering analyses. For the lower bound, a nonzero value of $ r_{\mathrm{min}}$ is often selected to suppress spurious low-$r$ features that arise from artifacts in the Fourier transform. In contrast, the physics-informed prior in our GP framework mitigates these distortions directly, removing the need for such \textit{ad hoc} corrections. Furthermore, it has been shown that the choice of the upper bound $R$ can significantly influence coordination number estimates~\cite{skinner_benchmark_2013}. Here, this source of uncertainty is explicitly accounted for by sampling from the posterior, enabling principled propagation of uncertainty into the coordination number distribution.

In all technicality, $n_{\alpha,\beta}$ is a functional operator that maps from some function space to $\mathbb{R}^+$. This motivates the use of an alternate notation $\tilde{\boldsymbol{n}}[g_{\alpha,\beta}(r);R]$ to emphasize this fact. Suppose now that you had two different RDF functions, $f(r)$ and $g(r)$, as well as two scalars $a$ and $b$. It is clear that,
\begin{align}
   \tilde{\boldsymbol{n}}[ag(r) + bf(r);R] = a\tilde{\boldsymbol{n}}[g(r);R]+ b\tilde{\boldsymbol{n}}[f(r);R]
\end{align}
which implies the number distribution of particles must also be a Gaussian process due to the linearity of the transformation. This feature implies we have the distribution, 
\begin{align}
    \tilde{\boldsymbol{n}}[g(r);R] \sim \mathcal{N}\left(m,s^2 \right) \label{eq:coord1}
\end{align}
where the mean $m$ is given by
\begin{align}
    & m =  \tilde{\boldsymbol{n}}[\mu_{\text{Post,RDF}};R]  = 4\pi\rho \int_0^R  \mu_{\text{Post,RDF}}r^2\ dr \\
    & =  4\pi\rho \int_0^R\left( \mu(r) + K(r,\boldsymbol{q})(\hat{\boldsymbol{K}}_{\boldsymbol{q},\boldsymbol{q}} + \omega^2\hat{\boldsymbol{I}})^{-1} (\boldsymbol{Y} - \boldsymbol{\mu}_{\boldsymbol{q}}) \right)r^2dr
\end{align}
and standard deviation $s$ is given by
\begin{align}
    & s^2 =\tilde{\boldsymbol{n}}[\tilde{\boldsymbol{n}}[ \Sigma_{\text{Post,RDF}};R];R] =  (4\pi\rho) ^2\int_0^R\int_0^R   \Sigma_{\text{Post,RDF}} \cdot (r 
    \ r')^2drdr' \\ 
    & =  (4\pi\rho)^2\int_0^R\int_0^R \left(K(r,r') - K(r,\boldsymbol{q})  (\hat{\boldsymbol{K}}_{\boldsymbol{q},\boldsymbol{q}} + \omega^2\hat{\boldsymbol{I}})^{-1} K(\boldsymbol{q},r')\right) \cdot (r\ r')^2 drdr'. \label{eq:coord2}
\end{align}
In principle these could be computed analytically if the form of $K$ and $\mu$ permitted, however we again opt for a numerical quadrature to not restrict the kernel design process outlined above. 

Although it might be tempting to use the distribution of $\tilde{\boldsymbol{n}}[g(r);R]$ near the first minimum of the RDF to estimate the first coordination number, this approach does not produce the correct result. Note the first coordination number is given by,
\begin{align}
    n_1 = \tilde{\boldsymbol{n}}[g(r);r_1],
\end{align}
where $r_1$ is the location of the first local minima in $g(r)$. Due the underlying RDF being a random function, its corresponding first minima will also be a random variable. This implies that $n_1$ should be a mixture of Gaussians. Not only that, it really is an infinite mixture of Gaussians as there is always a non-zero probability of the RDF having a maxima on the entire support of $g(r)$ due to our assumption that the process is Gaussian. This would mean we would need to first compute the probability distribution over $r_1$ first, and then propagate that uncertainty through into a computation of $n_1$.  

Unfortunately, if the structural prior is at all complicated then the estimation of the probabilities $p(r_1)$ or $p(n_1)$ may not even be analytical, clearly leading to difficulty. In this work, we resolve this issue using a Monte Carlo approach. For each sampled RDF, we identify the minimum $r_1^*$ via a search algorithm, then evaluate eq \eqref{eq:coord0} at $R =r_1^*$ to compute the corresponding coordination number. Repeating this procedure builds up a histogram of coordination numbers across the ensemble. A similar strategy may be taken to obtain any quantity of interest derived from $g(r)$ or $S(q)$. 

\section{Results and Discussion}

Having established the theoretical framework, we now demonstrate its utility on both simple and complex liquids through synthetic and experimental scattering data. (1) In a liquid argon scattering experiment, we demonstrate excellent agreement between GP-derived structure factors and results from a gold-standard neutron scattering analysis. Beyond numerical accuracy, the non-stationary GP provides enhanced physical interpretability through kernel heat maps and posterior covariance matrices, which visualize the relationship between momentum- and real-space features. (2) To validate the non-stationary GP framework in a molecular system, we applied the method to simulated liquid water with a known ground truth. We find that the GP reconstructs the real-space RDF, even under moderate noise. (3) Finally, we apply the framework to experimental X-ray scattering experiment of liquid water. Here, the GP yields a novel prediction for the oxygen-oxygen RDF with uncertainty quantification. The posterior is then propagated to estimate posterior predictive statistics for the first and second peak positions and heights, as well as the coordination number.

\subsection{Liquid Argon}

We begin by examining the quintessential neutron weighted argon structure factor measured by J. L. Yarnell \cite{yarnell_structure_1973}. Although the dataset incorporates \textit{post hoc} modifications to address multiple scattering, background scattering, finite sample volumes, and noise, it remains widely regarded as a benchmark dataset in the field. However, a key issue is that denoising alters the uncertainty estimation in the non-stationary GP method. To approximate the original, pre-denoised data and preserve realistic uncertainty estimates, we reintroduced constant Gaussian noise ($\sigma_{\text{noise}}^2 = 0.04$) to the input estimated from Figure 5 in Yarnell's manuscript. 

Now, for the non-stationary GP construction of liquid argon, which has no bonded contributions, the hyperparameter vector is reduced to,
\begin{align}
    \boldsymbol{\theta} = [r_0,s_0,\ell,\text{Max},\text{Slope},\text{Loc},\text{Decay},\omega]^T.
\end{align}

\begin{figure}
    \centering\includegraphics[width=1\linewidth]{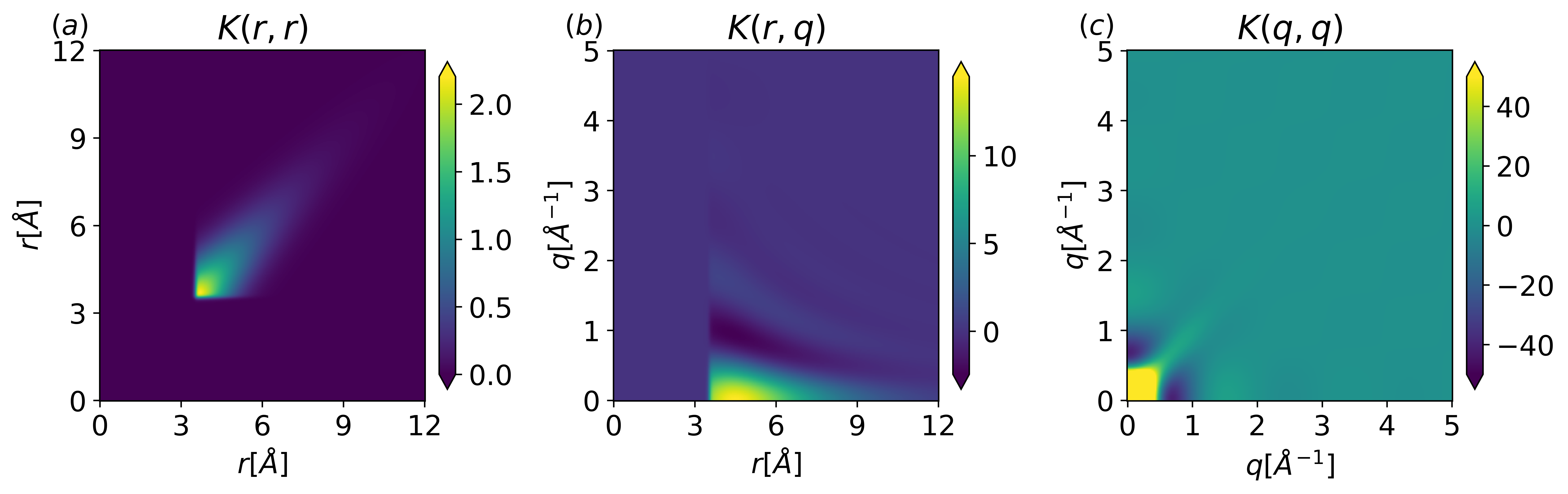}
    \caption{Gaussian process kernels after hyperparameter fitting of argon at temperature $T=85$[K] and density $\rho = 0.02125 \ [\text{atom}/\si{\angstrom}^3]$. Left corresponds to eq \eqref{eq:Krr}, middle corresponds to eq \eqref{eq:Krq}, and right corresponds to eq \eqref{eq:Kqq}. The colorbar represents the range of values indicated by the colormap. Any values outside the specified range are clipped and displayed using the colors corresponding to the nearest boundary.}
    \label{fig:kernelviz}
\end{figure}

Figure \ref{fig:kernelviz} visualizes eqs \eqref{eq:Krr} through \eqref{eq:Kqq} before hyperparameter optimization. Notably the kernel matrices exhibit the distinct structural characteristics enforced through the prior. Specifically, the lack of structure at low radius values in eqs \eqref{eq:Krr} and \eqref{eq:Krq} corresponds to the excluded volume of the argon atoms. The presence of this feature in the prior distribution suggests the atomic size is learned during the LMLH optimization of the prior mean hyperparameters rather than conditioning on the observed data. At medium to large values of $r$ there is a clear periodic structure in $q$, indicating both positive and negative correlations. This is an expected feature due to the underlying integration factor in eq \eqref{eq:radFT} being a decaying sinusoid. Lastly, notice the magnitudes involved in each correlation. While the maximum value in the $K(r,r)$ correlation is typical of Ar, the magnitudes in $K(q,q)$ are greater than what is observed in experiments. This results in an increased flexibility in the low $q$ region that is inconsistent with known limiting behaviors of $S(q)$. 

For example, the high variance at low $q$ ($\sigma_q^2 > 40$, Figure \ref{fig:kernelviz} (c) ) arises from model misspecification. This discrepancy appears to result from the absence of constraints on total density and isothermal compressibility. To understand this, consider the case $q=0$, where the sinc term in eq \eqref{eq:radFT} tends to 1 in the $q \to 0$ limit so that $S(0) = 1 + 4\pi\rho\int_0^\infty (g(r)-1) r^2  dr$. The behavior at $q = 0$ is then determined by the well-known compressibility equation, which relates $S(q)$ to the isothermal compressibility, 
\begin{align} 
    \lim_{q\rightarrow 0} S(q) = \rho k_B T \chi_T.
\end{align} 
Hence, the large prior variance at $q = 0$ indicates the function distribution does not have a fixed isothermal compressibility. Thermodynamic quantities of this type could be incorporated into the model directly as constraints, ensuring that all realizations of the experimental data are consistent with known thermodynamic quantities. Several strategies exist to enforce such consistency, including the use of warping functions, the design of kernel and mean functions that inherently satisfy the constraints, or the incorporation of auxiliary data to implicitly embed them \cite{swiler_survey_2020}. Although such constraints are not utilized here, it remains a promising direction for future exploration of GPs as mathematical models in liquid state theory. 

\begin{figure}
    \centering
    \includegraphics[width=0.85\linewidth]{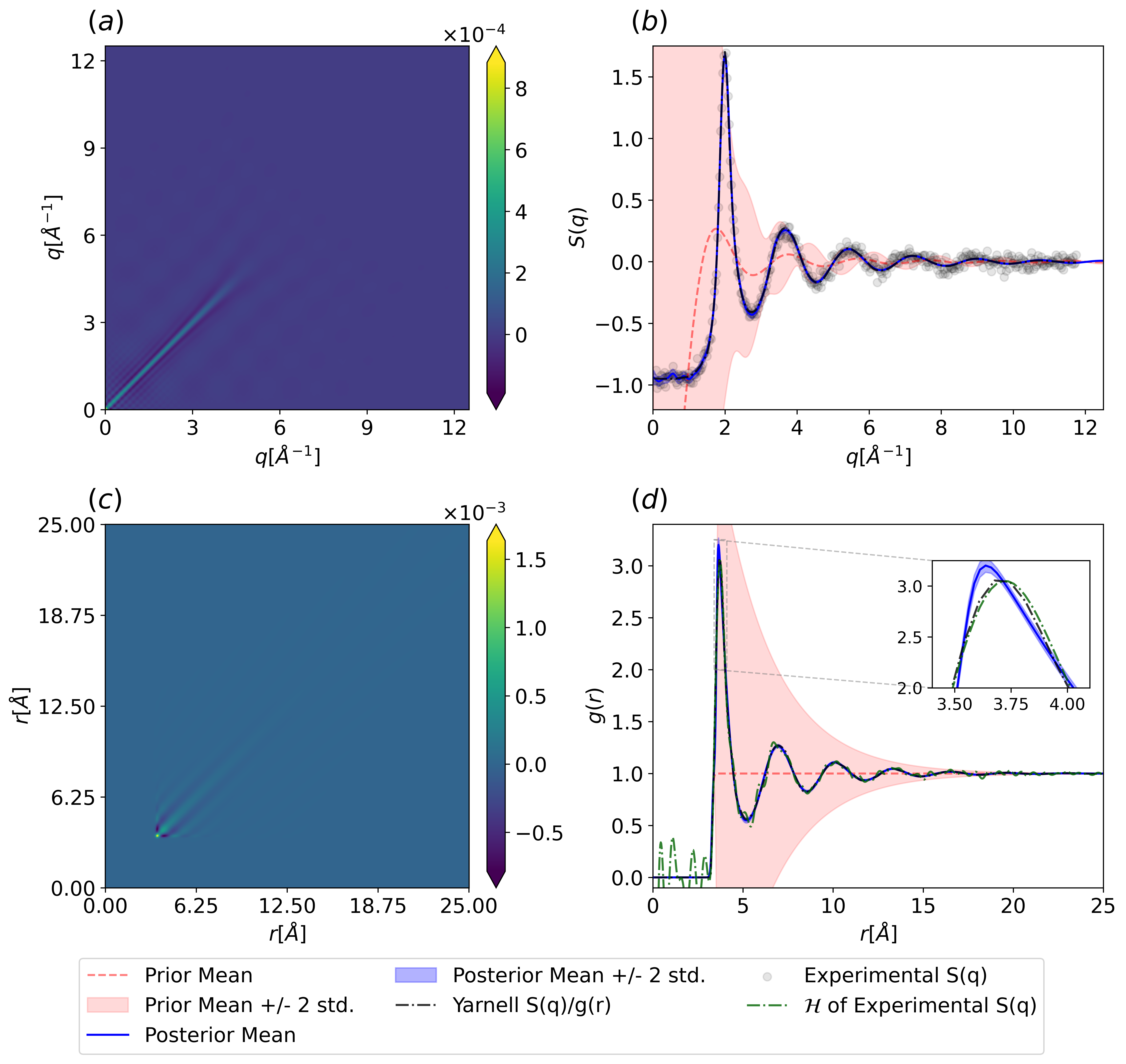}
    \caption{Posterior of the Gaussian process fit to argon at temperature $T=85$[K] and density $\rho = 0.02125 \ [\text{atom}/\si{\angstrom}^3]$. (a) Posterior covariance in $q$-space from eq \eqref{eq:SQ_post_sigma}. (b) Prior and posterior distribution for the argon structure factor from eq \eqref{eq:SQ_post_mu}. (c) Posterior covariance in $r$-space from eq \eqref{eq:RDF_post_sigma}. (d) Prior and posterior argon RDF from eq \eqref{eq:RDF_post_mu}. The colorbars are clipped similarly to Figure \ref{fig:kernelviz}.}
    \label{fig:argon_fig}
\end{figure}

With these kernels, we perform hyperparameter optimization by minimizing eq~\eqref{eq:LMLH}, after which the prior distribution over structure factors is conditioned on the data to yield a posterior distribution, along with the associated distribution over the RDF. This procedure is described by eqs \eqref{eq:post_sq} through \eqref{eq:SQ_post_sigma} and \eqref{eq:post_rdf} through \eqref{eq:RDF_post_sigma}. 
The posterior mean and covariance in $\boldsymbol{q}$ and $\boldsymbol{r}$ space are presented in Figure \ref{fig:argon_fig}.

The direct rFT of the data in Figure \ref{fig:argon_fig} (d) clearly exhibits $q_\text{max}$ cutoff errors, manifesting as high frequency oscillations. As commented by Lorch in 1969, these high frequency oscillations are often "erroneously identified as truncation ripples by other workers" \cite{lorch_neutron_1969}. In the Yarnell interpretation, truncation ripples were removed through an iterative procedure. First, the structure factor $S(q)$ was artificially extended to the compressibility limit ($q=0$), then directly Fourier-transformed to produce an initial estimate of $g(r)$. Next, $g(r)$ was set to zero in the low-$r$ region ($0 \leq r \leq 0.8d$, where $d$ is an estimated atomic diameter) and inverse Fourier-transformed back to yield an updated $S(q)$. This process was repeated iteratively until it was no longer necessary to set $g(r)$ to zero at low-$r$. This iterative procedure is well-established in neutron scattering analysis, and the non-stationary GP framework naturally preserves the spirit of this procedure. The process of optimizing eq \eqref{eq:LMLH} formally performs the same scheme. In practice, both Yarnell’s method and the non-stationary GP yield nearly identical predictions of the real-space structure, with deviations likely arising from the denoising procedure or imperfect hyperparameter optimization. To account for this type uncertainty in the GP formalism one would increase the hierarchy of the optimization and propagate $p(\boldsymbol{\theta}|\boldsymbol{Y)}$ into the $g(r)$ distribution. Due to the associated computational cost as well as the negligible difference to Yarnell's results we did not explore this avenue, however it provides a clear direction for future work.

The non-stationary GP methodology also provides us with direct access to the posterior covariance matrix in real-space. We can see in Figure \ref{fig:argon_fig} (c) that the posterior covariance in real-space exhibits a highly non-stationary structure that is fully consistent with the physical constraints imposed in the kernel design stage. Namely, the zero covariance at short-range exactly mimics the certain low-$r$ limit constraint, while the decaying covariance at high-$r$ reflects the decay of the RDF oscillations to unity. In momentum-space, the decreasing posterior covariance of the structure factor as a function of $q$ demonstrates that the physics-informed prior on the RDF naturally leads to a physically consistent estimate of $S(q)$ through the rFT (Figure \ref{fig:argon_fig} (a)).

\subsection{Water}

We now turn our attention to liquid water, a considerably more complex system due to the presence of chemical bonds and three partial structure factors. When inferring real-space structures in bonded systems, the non-stationary GP framework requires an additional prior mean term, given by eq \eqref{mean_term}, to handle the bonded part of the structure for the oxygen-hydrogen and hydrogen-hydrogen partial structure factors. Other than the additional hyperparameters introduced by the prior mean, the non-stationary GP regression proceeds in the exact same manner as in the liquid argon case.

\subsubsection{Simulated Liquid Water}

First, we analyzed an artificial noisy unweighted structure of simulated water obtained from the flexible TIP4P/2005f water model~\cite{gonzalez_flexible_2011} (for simulation details, see Supporting Information Section F). The goal of this test was to verify that the non-stationary GP accurately recovers the ground truth real-space structure from noisy momentum-space data. This validation step is essential, as it builds confidence in the methodology before applying it to experimental data where the ground truth real-space structure is unknown. 

\begin{figure}
    \centering
    \includegraphics[width=1\linewidth]{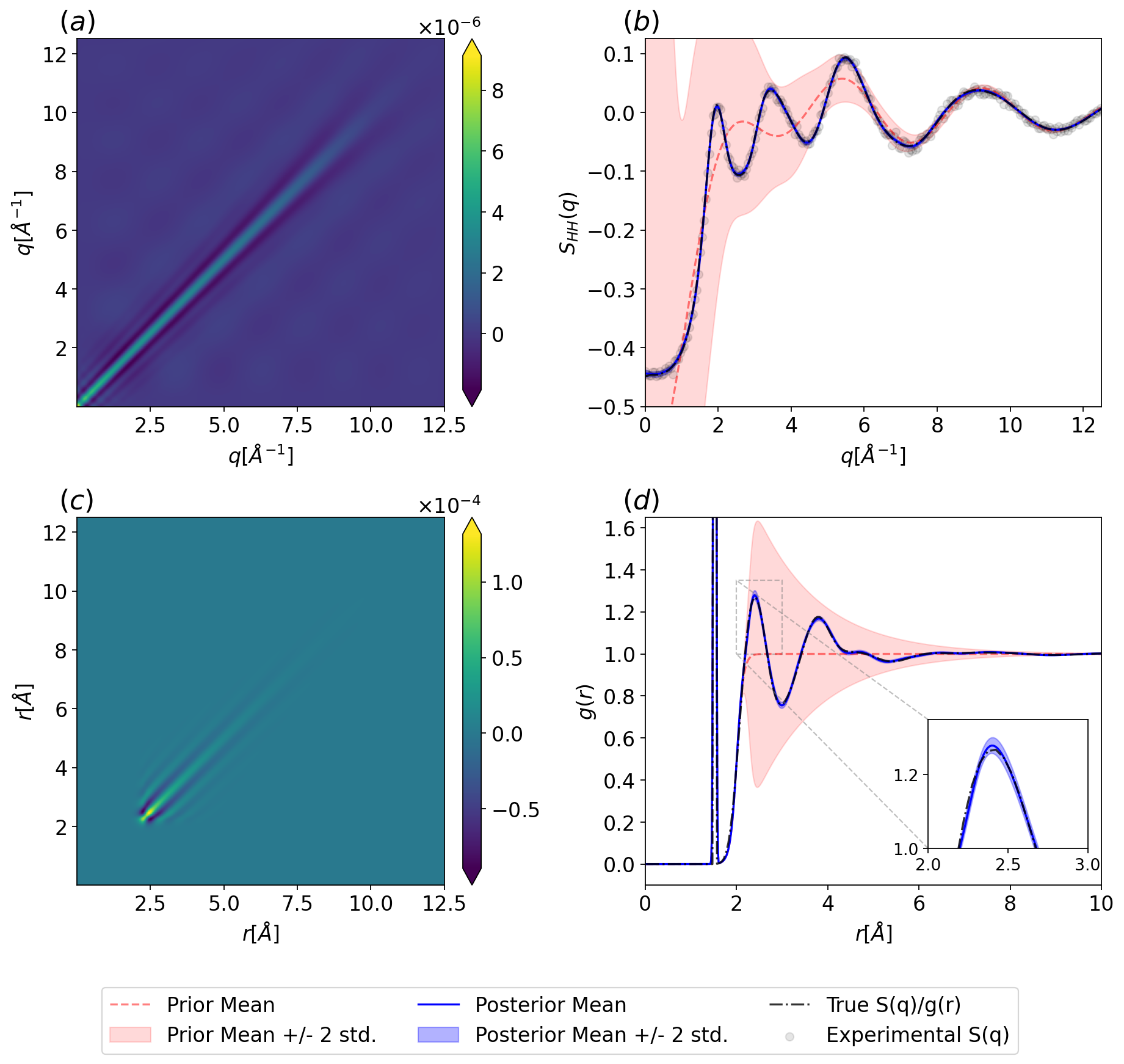}
    \caption{Posterior of the Gaussian process fit to structure factors derived from NVT simulations at a temperature of 298.15~K and density 1~g~cm$^{-3}$ with flexible TIP4P/2005f water. (a) Posterior covariance in $q$-space from eq \eqref{eq:SQ_post_sigma}. (b) Prior and posterior distribution for the hydrogen-hydrogen structure factor from eq \eqref{eq:SQ_post_mu}. (c) Posterior covariance in $r$-space from eq \eqref{eq:RDF_post_sigma}. (d) Prior and posterior hydrogen-hydrogen RDF from eq \eqref{eq:RDF_post_mu}. The sharp feature observed at $\sim$ 1.6 $\si{\angstrom}$ represents the distance between the hydrogen atoms in a water molecule.}
    \label{fig:sim_water_fig_HH}
\end{figure}

In Figure \ref{fig:sim_water_fig_HH}, we show the posterior covariance and hydrogen-hydrogen partial structure factor and RDF including the GP prior (red), posterior (blue), perturbed structure factor (black dots), and the ground truth (dashed black line). The hydrogen–hydrogen partial structure factor is presented here because it includes both bonded and non-bonded contributions, resulting in a more complex correlation structure with more hyperparameters than the oxygen–oxygen case, and thus poses a greater challenge for the method. The data in Figure \ref{fig:sim_water_fig_HH} (b) clearly shows that the hyperparameter learning and regression result in an excellent posterior representation of the hydrogen-hydrogen partial structure factor as well as its rFT to real-space in Figure \ref{fig:sim_water_fig_HH} (d). The bonded and non-bonded regions of the hydrogen–hydrogen partial RDF are well captured by the model with the ground truth lying within the estimated RDF posterior distribution. While we only show the hydrogen-hydrogen partial posterior here, we note that the oxygen-hydrogen and oxygen-oxygen posterior distributions exhibit strong qualitative agreement and are provided in Section F of the Supporting Information. 

\subsubsection{Experimental X-ray Scattering of Liquid Water}

We now turn to the analysis of a broadened X-ray scattering dataset for liquid water reported by Skinner and coworkers \cite{skinner_benchmark_2013}. In X-ray scattering experiments on water, the signal is dominated by oxygen-oxygen correlations due to the weak scattering cross section of hydrogen, arising from its single electron. As a result, assuming that background scattering corrections in the original dataset were appropriately handled, the non-stationary GP model in this case needs only to infer the oxygen-oxygen correlation. We then compared our predictions to those of Skinner’s interpretation. In their analysis, a variable Lorch modification function developed by Soper and Barney \cite{soper_extracting_2011} was applied with fixed parameters ($a = 2.8$ and $b = 0.5$ \AA) and error propagation was performed using the method of Weitkamp \cite{weitkamp_hydrogen_2000}.

\begin{figure}
    \centering
    \includegraphics[width=0.95\linewidth]{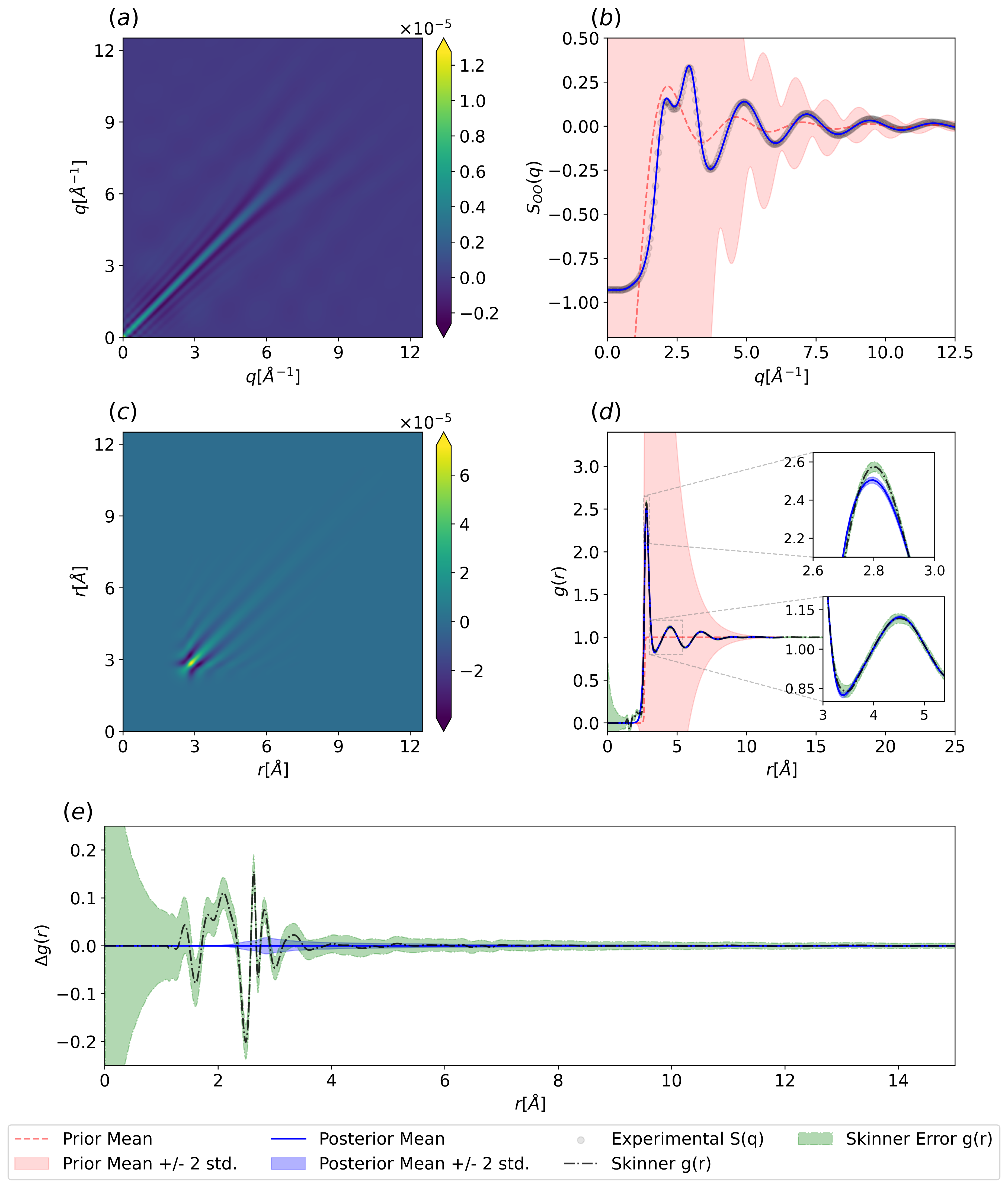}
    \caption{Posterior of the Gaussian process fit to the X-ray scattering data. (a) Posterior covariance in $q$-space from eq \eqref{eq:SQ_post_sigma}. (b) Prior and posterior distribution for the oxygen-oxygen structure factor from eq \eqref{eq:SQ_post_mu}. (c) Posterior covariance in $r$-space from eq \eqref{eq:RDF_post_sigma}. (d) Prior and posterior oxygen-oxygen RDF from eq \eqref{eq:RDF_post_mu}. (e) GP Mean subtracted comparison between the uncertainty estimates from the non-stationary GP approach and Skinner's interpretation\cite{skinner_benchmark_2013}.}
    \label{fig:exp_water}
\end{figure}

The non-stationary GP applied to the structure factor provides an excellent representation of the underlying structure given the noisy experimental X-ray scattering data. More insightful, however, is the comparison between the non-stationary GP model and Skinner's interpretation shown in Figure \ref{fig:exp_water}(d)-(e). While the mean predictions from both methods closely align at and beyond the first peak, significant differences emerge at lower distances. Specifically, Skinner's interpretation exhibits non-physical oscillations below the collision diameter of the oxygen atom, an artifact of Fourier truncation, that leads to non-physical negative values of $g(r)$. Even more striking are the differences in uncertainty estimates between the two methods. Skinner's uncertainty rapidly increases at low-$r$, primarily reflecting known limitations of the applied error-estimation procedure \cite{weitkamp_hydrogen_2000}, making it unclear whether these uncertainties genuinely represent data-informed variability or merely methodological artifacts. Conversely, the non-stationary GP uncertainty profile matches physical expectations: negligible uncertainty at low-$r$, a pronounced increase reaching a maximum around the first peak, followed by a gradual decay to negligible uncertainty at large distances. Notably, our interpretation predicts a slightly larger uncertainty in the local structure of water in the first solvation shell.

\begin{figure}
    \centering
    \includegraphics[width=1\linewidth, trim=0 0 0 125, clip]{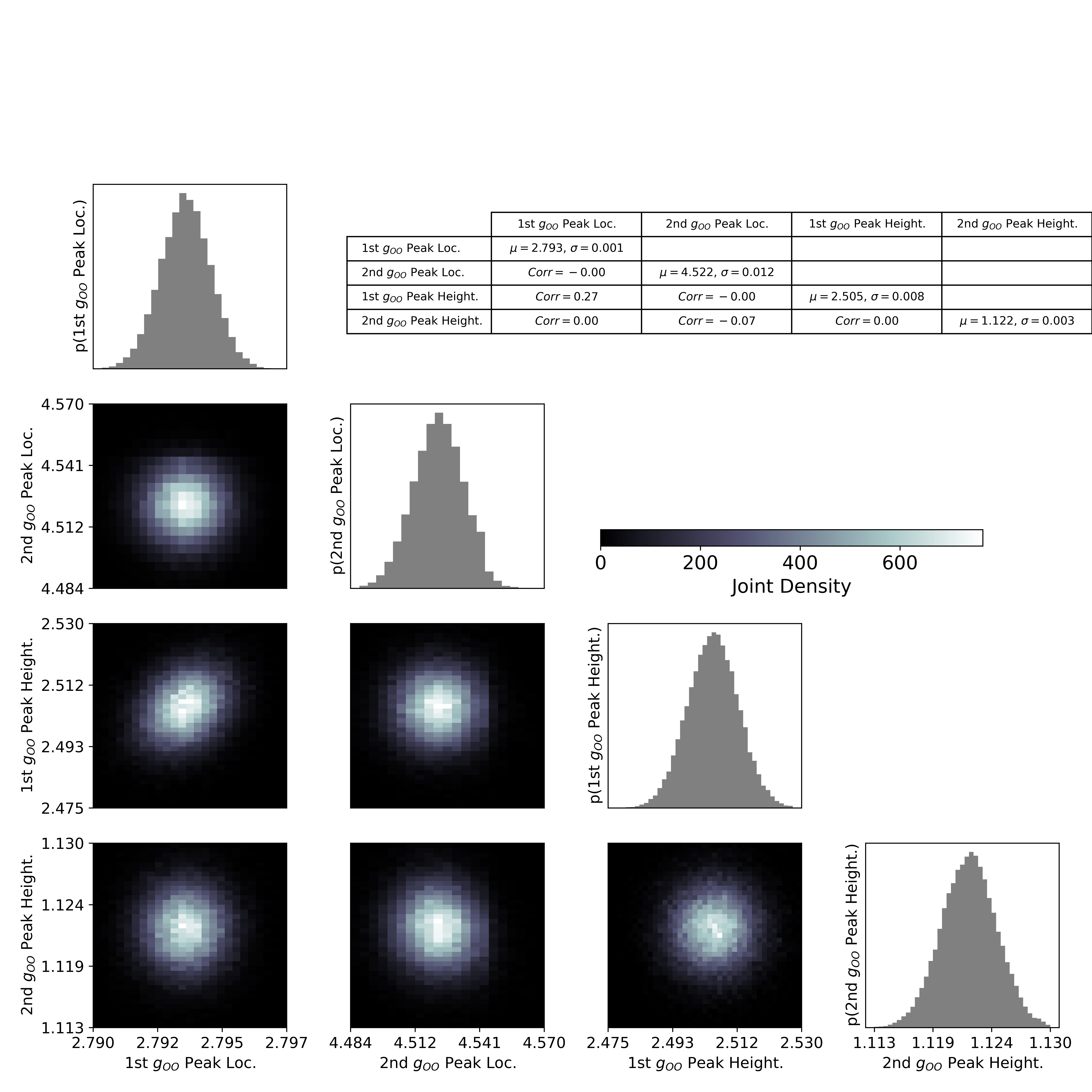}
    \caption{This corner plot shows the joint distribution of the 1st and 2nd RDF peak locations and heights. The table shows the mean and standard deviations values of the marginals along its diagonal. The off diagonal terms are the Pearson correlation coefficients in the joint marginal distributions.}
    \label{fig:peaks}
\end{figure}

\begin{figure}
    \centering
    \includegraphics[width=1\linewidth, trim=0 0 0 0, clip]{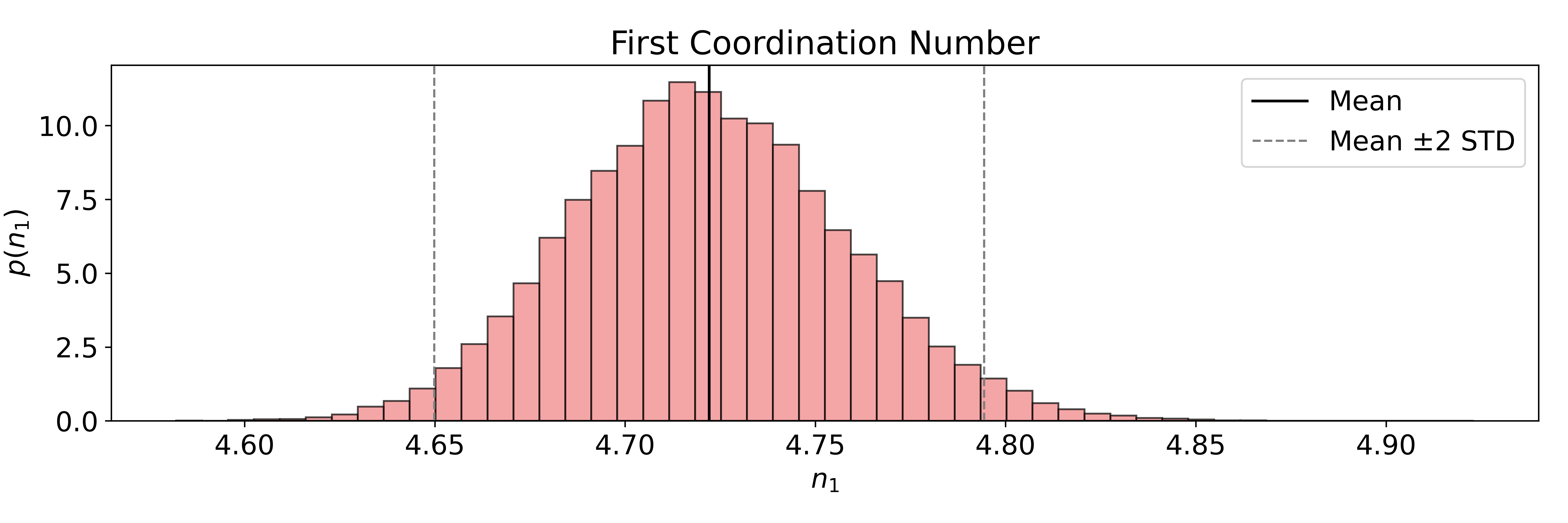}
    \caption{Histogram estimation of the coordination number probability density derived from samples of the non-stationary GP posterior.}
    \label{fig:coord}
\end{figure}

A physically-justified posterior distribution on the RDF can subsequently be used to estimate statistics on other observables such as peak heights and peak positions (Figure \ref{fig:peaks}) and coordination number (Figure \ref{fig:coord}). Since the distributions are nearly Gaussian, we present the mean plus or minus two standard deviations ($\mu \pm 2 \sigma$) as a summary statistic. The first peak location is estimated to be 2.793 $\pm$ 0.002, and first peak height is 2.505 $\pm$ 0.016. The joint 2D marginal distributions over peak location and peak height show near zero correlation for every two parameter set, aside from a slight positive correlation (0.27) between the first peak height and first peak location. Finally, the estimated first coordination number is 4.722 $\pm$ 0.07, which is in good agreement with the generally accepted value determined from X-ray scattering data of 4.7 \cite{head-gordon_tetrahedral_2006}.

Given that water is of central importance for a wide variety of fields and the subject of substantial investigation over its local structure, it is worth reflecting on which structural interpretations should serve as benchmarks for molecular modeling. The non-stationary GP framework presented here offers not only a physics-informed reconstruction of the RDF, but also a posterior predictive distribution that quantifies uncertainty in a principled way. This makes it particularly well-suited for benchmarking simulations, where one expects model predictions, in this case RDFs, coordination numbers, peak heights, and positions, to fall within credible intervals that reflect both experimental noise and structural ambiguity. While no interpretation is free from assumptions or limitations, the GP-based approach provides a transparent and reproducible framework that balances physics with statistical rigor, and is likely to be a more robust choice for comparison against molecular models.

\section{Discussion}

The non-stationary GP framework offers a principled, data-driven approach to structural inference by combining Bayesian inference with physics-informed priors. We demonstrate accurate fits for test systems ranging from liquid Ar, TIP4P/2005f water, to experimental X-ray scattering of liquid water with only modest physical assumptions. The method operates on a continuous domain in both momentum- and real-space to mitigate problems with binning artifacts or $q_{max}$ truncation errors. The model effectively filters normal random noise present in the experimental observations, which ultimately stabilizes any subsequent analysis of downstream properties such as peak heights, locations, and coordination numbers. By avoiding simulation-specific biases (e.g., from force field parameters, thermostats, or numerical integrators), the GP posterior provides a reproducible and physically grounded benchmark. Moreover, its flexibility allows for the inclusion of additional physics-based constraints (e.g., isothermal compressibility limits, virial equations, Kirkwood-Buff integrals \cite{kirkwood_statistical_1951}), making it a powerful tool for bridging scattering data with macroscopic thermodynamic properties.

The non-stationary GP method also hints at a deeper connection between structure and interatomic potentials when there is noise present in the data. The Henderson inverse theorem, which shows that the radial distribution function for pairwise additive and homogeneous systems has a pair potential that is unique up to an additive constant, is derived in the noiseless limit \cite{henderson_uniqueness_1974}. However, as shown empirically by Soper, when noise is present there is an ensemble of potential energy functions corresponding to the scattering data target \cite{soper_tests_2001}. We hypothesize that the non-stationary GP method can be considered a Laplace approximation on the true structural posterior determined by such a potential energy ensemble. Adopting this philosophical perspective could significantly enhance our understanding of structure–thermodynamics relationships and improve the accuracy of thermodynamic predictions.

Despite these advantages, several limitations remain. The nature of scattering experiments themselves pose the problem that species with low scattering length densities may be effectively invisible in the total signal. Additionally, all structure-analysis methods confront the underdetermined nature of the Faber–Ziman decomposition for multicomponent systems and mixtures, which permits multiple RDF solutions consistent with the same total scattering data leading to non-uniqueness \cite{soper_uniqueness_2007}. The GP framework could be extended to these cases by representing each partial structure factor as a linear combination of non-stationary GPs; however, this approach quickly increases the number of hyperparameters to be inferred, leading to higher computational cost. It may also be possible to develop methods analogous to EPSR that consistently integrate experimental data with molecular models of the interatomic potential within a Bayesian framework (c.f. \cite{shanks_transferable_2022}), though, to our knowledge, no such algorithm has been attempted.

Furthermore, absent \emph{perfect} experimental data processing procedures (e.g., background, multiple scattering, inelasticity, etc.) and scattering statistics resulting from an infinite radiation flux, no fluid structure interpretation can be entirely unbiased. While our approach assumes these corrections are accurate, extending the GP framework to model them explicitly would be a valuable step toward a more complete and uncertainty-aware analysis. In principle, the GP framework can incorporate such systematic corrections hierarchically, for example by using non-Gaussian likelihoods for scattering corrections, performing Bayesian inference over parametric models of systematic errors, and estimating time-of-flight uncertainty via error-in-variables approaches. 

Finally, the GP prior mean and kernel used here represent just one of many choices that can satisfy the physical constraints. Future work could explore alternative priors that more tightly enforce thermodynamic behavior or integrate knowledge of interatomic potentials obtained through other Bayesian schemes \cite{shanks_transferable_2022,shanks_accelerated_2024}. As the methodology evolves, refining and exploring alternative priors is likely to improve interpretability, computational efficiency, and predictive accuracy.

\section{Conclusions}

We introduce a method for rigorous uncertainty quantification and propagation in experimentally derived radial distribution functions using physics-informed, non-stationary Gaussian process regression. This approach constructs a minimal yet physically expressive kernel that preserves the Fourier duality between the structure factor and the radial distribution function. By addressing pervasive challenges in the Fourier transformation of momentum-space scattering data and incorporating Bayesian inference, our approach offers a robust and interpretable alternative to traditional structural analysis methods. 

Applied to both simple and complex liquids, the model yields physically reasonable posterior distributions for RDFs that capture both mean behavior and structural uncertainty. Crucially, the non-stationary GP framework achieves this without relying on  computationally intensive molecular simulations that may be affected by systematic model bias imposed by force field assumptions. Its flexibility allows for principled incorporation of physical knowledge and integration of data preprocessing steps through hierarchical modeling. Taken together, we conclude that the Bayesian framework established in this study may represent the best path towards an unbiased assessment of fluid structure.

\newpage

\section{Acknowledgments}

This study is supported by the EFRC-MUSE, an Energy Frontier Research Center funded by the U.S. Department of Energy, Office of Science, Basic Energy Sciences under Award No. DE-SC0019285. The support and resources from the Center for High Performance Computing at the University of Utah are gratefully acknowledged. We would like to thank Aryan Deshwal for reading a preliminary version of the manuscript and providing helpful comments on Gaussian processes and kernel design.

\section{Author Contributions}

H.W. Sullivan: conceptualization (equal), algorithm development (lead), code implementation (lead), writing - original draft (equal) B.L. Shanks: conceptualization (equal), algorithm development (supporting), writing - original draft (equal) M.P. Hoepfner: algorithm development (supporting), writing - review and editing (lead), funding acquisition (lead) 

\section{Data Availability}

Structure factors, radial distribution functions with credibility intervals, and non-stationary GP hyperparameters are available from the corresponding author upon reasonable request and also provided on Github at \url{https://github.com/hoepfnergroup/LiquidStructureGP-Sullivan}

\section{Supporting Information}
\appendix
\section{Notation}

\begin{itemize}
    \item A symbol $f$ is a scalar. 
    \item A bold symbol $\boldsymbol{f}$ is a column vector. 
    \item A bold symbol with a hat $\hat{\boldsymbol{f}}$ is a matrix.
    \item A bold symbol with a tilde $\Tilde{\boldsymbol{f}}$ is an operator that acts on  functions. The arguments of an operator are put between $[\ \ ]$, arguments to the right of a semicolon $;$ indicate implicit arguments of the operator. Implicit arguments are not exhaustive and are purely shown for pedagogy.
    \item An operator with a subscript $\Tilde{\boldsymbol{f}}_x$ is meant to indicate that it is treating the argument as a function of the subscript alone. For example $\Tilde{\boldsymbol{f}}_x[g(x,y,z)]$ is the result of acting with $\Tilde{\boldsymbol{f}}$ on the function $x \mapsto g(x,y,z)$ which has implicit dependence on y and z.
    \item A numerical approximation to an operator is indicated with a blackboard bold $\mathbb{f}$. While the notation makes it look as if it acts on functions it is implied that it acts on discretized evaluations of the them.
    \item The abuse of notation $f(\boldsymbol{x})$ just means $[f(x_1),\dots,f(x_n)]^T$.
    \item $\boldsymbol{f} \sim p(\boldsymbol{\zeta}_1,\boldsymbol{\zeta}_2,\dots)$ means the column vector $\boldsymbol{f}$ is distributed by the p.d.f. $p$ parametrized by $\boldsymbol{\zeta}_1,\boldsymbol{\zeta}_2,\dots$.
    \item $\operatorname{cov}\left(f(x),g(y)\right)$ is the covariance between the function values $f(x)$ and $g(y)$.
    \item $f \sim \mathcal{GP}(\mu,K)$ means the function $f(x)$ is a Gaussian process whose finite index set $\boldsymbol{x}$ has function values distributed as a multivariate Gaussian with mean $\mu(\boldsymbol{x})$ and covariance $K(\boldsymbol{x},\boldsymbol{x})$.
    \item $\mathbb{R}^d$ is the whole real line in $d$ dimensions. If there is no superscript then it is one dimensional.
    \item $\mathbb{R}^{d,+}$ is the positive real line in $d$ dimensions. If there is no $d$ in the superscript then it is one dimensional.
    \item An integral without bounds implies it is taken over the whole support of the distribution.
\end{itemize}

\section{Mitigation of Numerical Errors} 

Although the GP formalism is well founded, it is still subject to the same constraints as any high-dimensional linear algebra program. The key issues we are concerned with include (but are not limited to) floating point errors, approximate sparsity of the kernel, unbounded derivatives, near-positive definiteness of matrices, and convergence of eigenvalues and eigenvectors. Often these aspects are left behind the curtain, only for the users to discover and solve on their own. Rather than following a \textit{do-it-yourself} approach, we are including these details to avoid potential issues in future implementations of the GP paradigm.

Kernel functions, and their corresponding kernel matrices, must be positive semi-definite (PSD) to ensure that they represent valid covariance matrices. Suppose we had a kernel function that is not PSD. This would imply the variance in the direction of at least one eigenvector would be negative, which is clearly nonsense probabilistically speaking. The approximate sparsity of our kernel, with many elements close to zero, may lead to a non-PSD matrix. To address this, we iteratively adjust the eigenvalues of non-PSD kernels. To do this first compute the eigenvectors $\boldsymbol{v}_i$ and eigenvalues $\lambda_i$ of the symmetrized kernel matrix.
\begin{align}
    \text{eigh}\bigg(\frac{\hat{\boldsymbol{K}}+\hat{\boldsymbol{K}}^T}{2}\bigg) = \{\lambda_i,\boldsymbol{v}_i\}
\end{align}
Then, we adjust each eigenvalue by adding the negative of the minimum eigenvalue plus a small offset, $\varepsilon$.
\begin{align}
    \lambda_{i,\text{Reconstruction}} = \lambda_i  - \min(\lambda_i) + \varepsilon
\end{align}
We can then reconstruct the matrix from the adjusted eigenvalues and eigenvectors. If the result is still not PSD, we repeat the reconstruction with a slightly larger $\varepsilon$. Effectively this is a post-hoc shrinkage method for covariance estimation. This adjustment introduces additive, uncorrelated, normally distributed noise in the GP samples.

However, there is no free lunch. Care must be taken during this step; if the underlying kernel function (and its associated hyperparameters) pathologically produces nearly singular or numerically non-PSD matrices, the $\varepsilon$ jitter term may grow large. This becomes increasingly relevant if the jitter term approaches the magnitude of the experimental noise $\omega$, which can compromise the reliability of the uncertainty quantification. Additionally, computing the log marginal likelihood now depends heavily on the eigenvalue decomposition of the kernel matrix. This reliance introduces sensitivity to degenerate eigenvalues, which can cause unbounded derivatives with respect to the hyperparameters. Such unbounded behavior leads to instability in the optimization process, where large or erratic gradients may cause oscillations or divergence. We opt to mitigate these derivative issues by bypassing the gradient with respect to the post-hoc shrinkage all together. By overwriting the automatic differentiation with the identity we can prevent any unboundedness arising from degeneracy. Unboundedness can also result from repeatedly applying exponential functions in both the mean and kernel. To address this, it’s useful to apply clamping functions to the inputs before taking the exponential, which prevents NaNs and infinite derivatives from disrupting optimization without significantly affecting the typical output.

With these subtleties addressed, we also encounter the issue of non-physical hyperparameters arising from the optimization process. Traditionally, this is managed by setting the prior to assign zero probability to non-physical values, thereby avoiding their selection. However, this approach conflicts with the typical structure of stochastic optimizers, which operate over parameter vectors in \(\mathbb{R}^d\). To work around this, we redefine each hyperparameter in terms of an unbounded "raw" parameter, optimizing in the raw space instead. Specifically, we apply a scaled sigmoid transformation and its logit inverse for each hyperparameter, such that forward and inverse transformations are given by
\begin{align}
    \boldsymbol{\theta}_i = \frac{u_i - l_i}{1+\exp(-\boldsymbol{\theta}_{i,\text{Raw}})} + l_i,\quad\quad \boldsymbol{\theta}_{i,\text{Raw}} = \ln \frac{\frac{\boldsymbol{\theta}_i - l_i}{u_i - l_i}}{1-\frac{\boldsymbol{\theta}_i - l_i}{u_i - l_i}}
\end{align}
where \(u_i\) and \(l_i\) are the upper and lower bounds, and \(\boldsymbol{\theta}_{i,\text{Raw}}\) is the unbounded parameter optimized with automatic differentiation. The last point we wish to make in this section is to note the requirement of Cholesky decomposition. One should never use a standard inversion algorithm on a kernel function. Paraphrasing Philipp Hennig, the use of a standard matrix inversion when computing a GP posterior should be considered a bug. The numerical instability introduced by such a procedure is immense and should be avoided. The positive definiteness of kernel matrices allows us the luxury of a Cholesky decomposition for the computation of the posterior instead. In all transparency, the application of the eigenvalue decomposition above may introduce similar errors to the standard matrix inversion we are avoiding with Cholesky decomposition.

In practice, we have found that deviations from the ground truth often occur in regions near the first RDF peak or the transition from bonded to non-bonded structures. This effect is due to insufficient hyperparameter tuning, such as improper learning rates or early stopping. Not learning these sharp uptick regions precisely may seem unimportant when the quality-of-fit for the majority of the RDF is excellent, however; our prior work has demonstrated that this region can inform predictions of the repulsive exponent in a ($\lambda$-6) Mie potential \cite{shanks_transferable_2022}. This region is also important for liquid water, where nuclear quantum effects (NQEs) involving the light hydrogen atoms have been shown to significantly broaden structural features in the OH and HH partial RDFs \cite{cheng_nuclear_2016}.

\section{Derivation of the rFT Operator}\label{sec:ONST}

Each of the partial structure factors $S_{\alpha,\beta}(\boldsymbol{q})$is related to the RDF $g_{\alpha,\beta}(\boldsymbol{r})$ via a standard 3D inverse Fourier transform. 
\begin{equation}
     S_{\alpha,\beta}(\boldsymbol{q}) -1 =\rho\int (g_{\alpha,\beta}(\boldsymbol{r})-1)\exp(i\boldsymbol{q}\cdot\boldsymbol{r})d\boldsymbol{r}
\end{equation}
If the material is isotropic then we may write this in terms of $r = |\boldsymbol{r}|$ and $q = |\boldsymbol{q}|$. To do this first consider a spherical polar co-ordinate system in real space where the polar angle points in the direction of $\boldsymbol{q}$. We can choose $\theta$ to be the angle between these two vectors. 
\begin{equation}
     = 2\pi\rho\int_{r=0}^{\infty}\int_{\theta=0}^\pi (g_{\alpha,\beta}(r)-1)\exp(iqr\cos(\theta))r^2 \sin(\theta) drd\theta
\end{equation}
Performing the integral over theta will give:
\begin{equation}
      S_{\alpha,\beta}(q) - 1 =  4\pi\rho\int_{r=0}^{\infty} (g_{\alpha,\beta}(r) -1)\frac{\sin(qr)}{qr}r^2 dr
\end{equation}
This implies the 3D Fourier transform of the $S(q)$ will give the $g(r)$
\begin{equation}
    g_{\alpha,\beta}(r) - 1 =   \frac{1}{2\pi^2\rho}\int_{q=0}^{\infty} (S_{\alpha,\beta}(q)-1)\frac{\sin(qr)}{qr}q^2 dq
\end{equation}
The difference in prefactor is due to the forward transform carrying the $\frac{1}{(2\pi)^3}$. This can be written in terms of  the radial Fourier transform (rFT) operator
\begin{align}
    \Tilde{\boldsymbol{\mathcal{H}}}_q[f(q)] = \frac{1}{2\pi^2\rho}\int_0^\infty f(q) \frac{\sin(qr)}{qr} q^2 dq, \quad \Tilde{\boldsymbol{\mathcal{H}}}^{-1}_r[f(r)] = 4\pi\rho\int_0^\infty f(r) \frac{\sin(qr)}{qr} r^2 dr.
\end{align}
The RDF structure factor relationship is then written succinctly as
\begin{align}
     S_{\alpha,\beta}(q) = 1+\Tilde{\boldsymbol{\mathcal{H}}}^{-1}_r[g_{\alpha,\beta}(r)-1],\quad g_{\alpha,\beta}(r) = 1 + \Tilde{\boldsymbol{\mathcal{H}}}_q[S_{\alpha,\beta}(q)-1].
\end{align}

\section{Visualizing Error Bars in the Posterior Distribution}

The posterior function can be visualized in several ways, the most intuitive being the use of error bars around the mean value. However, care must be taken in interpreting these error bars, as they reflect distinct sources of variation in the posterior at a given point $q$ or $r$. 

There are two choices we can make for the error bars, both of which stem from an underlying choice of $\boldsymbol{\Sigma}_{\text{Post}}$. The first corresponds to a \textit{noise-free} posterior, as outlined in eqs \eqref{eq:post_sq} through \eqref{eq:SQ_post_sigma}. While one might expect the standard deviation, $\boldsymbol{\sigma} = \text{diag}(\boldsymbol{\Sigma}_{\text{Post}})^{0.5}$, to satisfy the typical 68-95-99.7 rule with respect to the experimental data, this is not the case here. Since we are considering a \textit{noise-free} posterior, the visualization does not account for statistical fluctuations arising from $\omega$. Although the visualization does not reflect the statistical noise in the data, the underlying calculation accounts for it when conditioning on the observed data. This effect is seen easily by noticing the $\omega$ dependence in eqs \eqref{eq:SQ_post_mu} and \eqref{eq:SQ_post_sigma}.

For this work, we have chosen to present the \textit{noise-free} posterior, as a real-space counterpart to $\omega$ is not available. With the sources of variation established, the error bars on the posterior distribution are given by,
\begin{align}
    & \boldsymbol{\sigma}_q = \text{diag}(\boldsymbol{\Sigma}_{\text{Post}})^{0.5}, \tab \boldsymbol{\sigma}_r = \text{diag}(\boldsymbol{\Sigma}_{\text{Post,RDF}})^{0.5}.
\end{align}

\section{Experimental Argon Analysis Details}

The $S(q)$ dataset was generated via interpolation of the reported data in Yarnell 1973 \cite{yarnell_structure_1973}. Once the dataset was interpolated we added a zero mean normal noise with standard deviation $0.04$ to represent approximate error due to reactor source measurements as reported in Figure 5 in the original publication. 

\begin{figure}
    \centering
    \includegraphics[width=1\linewidth]{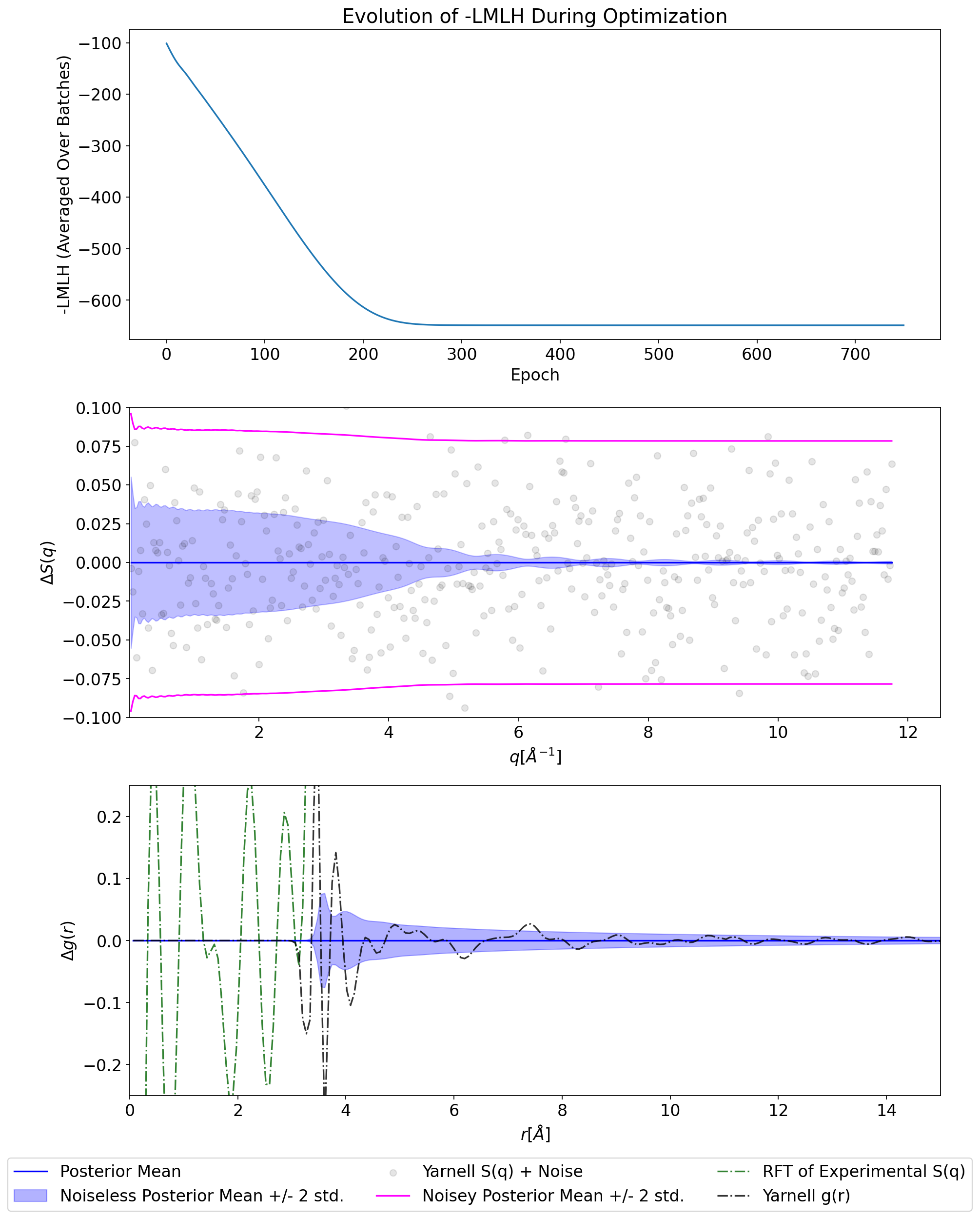}
    \caption{Top: The negative log marginal likelihood as a function of epoch during gradient descent optimization of equation \ref{eq:opt_target}. Middle: Difference between the non-stationary GP mean and the data with noiseless and noisy posterior credibility intervals. Bottom: Argon RDF residuals for the non-stationary GP (blue), Yarnell data with direct radial Fourier Transform (green dashed line) and Yarnell data with data smoothing and regularization (black dashed line).}
    \label{fig:LMLH_evolution_Ar}
\end{figure}

The set of parameters in table \ref{tab:Ar_init_opt_params} were optimized via equation \ref{eq:opt_target} using AdamW with a batch size of 400 over the 400 available data points and a learning rate of $1\cdot10^{-2}$. The optimization was run for 50 epochs, with a real space integration grid spanning from a small value close to zero (0.0001) to 25, divided into 1000 evenly spaced points. The evolution of the negative log marginal Likelihood (LMLH) during the optimization process is shown in figure \ref{fig:LMLH_evolution_Ar}. 

\begin{table}[H]
    \centering
    \begin{tabular}{|c|c|c|c|c|c|c|}
        \hline
        Parameter & Initial Value & Lower Bound & Upper Bound & Optimized Value & $\Delta$ & \% Change \\
        \hline
        $\ell$       & 1.300000 & 0.100000 & 2.500000 & 0.970872 & $-0.329128$ & $-13.71\%$ \\
        Max          & 0.800000 & 0.200000 & 4.000000 & 1.591555 & \phantom{$-$}0.791555 & \phantom{$-$}20.83\% \\
        Slope        & 25.050000 & 0.100000 & 50.000000 & 26.100749 & \phantom{$-$}1.050749 & \phantom{$-$}2.11\% \\
        Loc          & 3.250000 & 0.500000 & 6.000000 & 3.534076 & \phantom{$-$}0.284076 & \phantom{$-$}5.17\% \\
        Decay        & 0.525000 & 0.050000 & 1.000000 & 0.341316 & $-0.183684$ & $-19.34\%$ \\
        $\omega$   & 0.250050 & 0.000100 & 0.500000 & 0.039182 & $-0.210868$ & $-42.18\%$ \\
        $r_0$        & 3.250000 & 0.500000 & 6.000000 & 3.273637 & \phantom{$-$}0.023637 & \phantom{$-$}0.43\% \\
        $s$          & 25.050000 & 0.100000 & 50.000000 & 26.611479 & \phantom{$-$}1.561479 & \phantom{$-$}3.13\% \\
        \hline
    \end{tabular}
    \caption{Initial parameters, optimized parameters, and their bounds for experimental argon dataset.}
    \label{tab:Ar_init_opt_params}
\end{table}

\section{Simulated Water Data Analysis}

Molecular dynamics simulations of water were conducted using GROMACS 2023.3~\cite{abraham_gromacs_2015}, employing the flexible TIP4P/2005f water model~\cite{gonzalez_flexible_2011}. The system consisted of 1500 water molecules placed in a cubic periodic simulation box at the experimental density of approximately 1~g~cm$^{-3}$. Initial energy minimization was performed using the steepest-descent algorithm with a step size of 0.01~nm and a convergence criterion of 100~kJ~mol$^{-1}$~nm$^{-1}$. After minimization, the system underwent a brief 0.1~ns equilibration in the NVT ensemble at 298.15~K using the velocity-rescale thermostat~\cite{bussi_canonical_2007} with a coupling constant of 1~ps. Due to the flexibility of intramolecular potentials, the simulation timestep was set to 0.2~fs. Subsequently, the system was equilibrated for an additional 1~ns under NPT conditions at 298.15~K and 1~bar. The temperature was maintained at 298.15~K using the velocity-rescale thermostat with a coupling constant of 1~ps, while pressure was regulated with an isotropic C-rescale barostat~\cite{bernetti_pressure_2020} employing a coupling constant of 3~ps. Electrostatic interactions were computed via the particle mesh Ewald (PME) method~\cite{essmann_smooth_1995} with a cutoff of 1.0~nm, Fourier grid spacing of 0.10~nm, and fourth-order spline interpolation. Lennard-Jones interactions used a potential-shift cutoff scheme at 1.0~nm along with long-range dispersion corrections. Neighbor lists were updated every five simulation steps with the Verlet scheme, applying a buffer tolerance of $5\times10^{-3}$~kJ~mol$^{-1}$. Production simulation was carried out for 10~ns using the same settings as described for the NPT equilibration run. Intramolecular O-H bonds and H-O-H angles were explicitly flexible according to the TIP4P/2005f potential, and massless dummy site geometries were constrained using the LINCS algorithm of sixth order with one iteration~\cite{hess_lincs_1997}. Calculations of radial distribution functions were conducted using the Visual Molecular Dynamics (VMD) software~\cite{humphrey_vmd_1996}.

\begin{figure}
    \centering
    \includegraphics[width=1\linewidth]{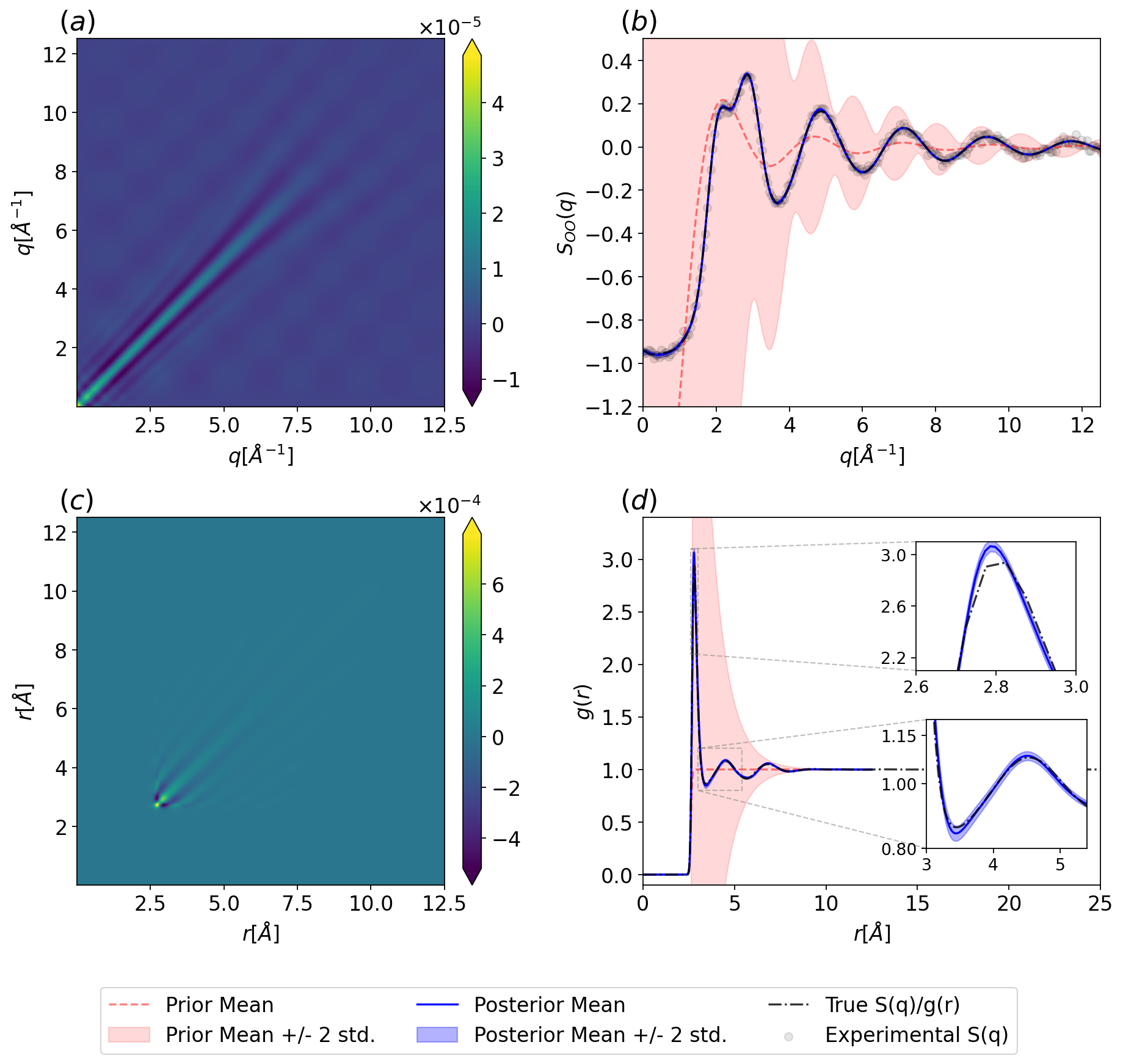}
    \caption{Posterior of the Gaussian process fit to structure factors derived from flexible TIP4P/2005f water for the oxygen-oxygen correlation. Top left corresponds to eq \eqref{eq:SQ_post_sigma}. Top right corresponds to eq \eqref{eq:RDF_post_sigma}. Bottom left corresponds to eq \eqref{eq:SQ_post_mu} and the diagonal of the covariance shown in the top left. Bottom right corresponds to eq \eqref{eq:RDF_post_mu} and the diagonal of the covariance shown in the top right. The colorbars are clipped similarly to Figure \ref{fig:kernelviz}.}
    \label{fig:sim_water_fig_OO}
\end{figure}

\begin{figure}
    \centering
    \includegraphics[width=1\linewidth]{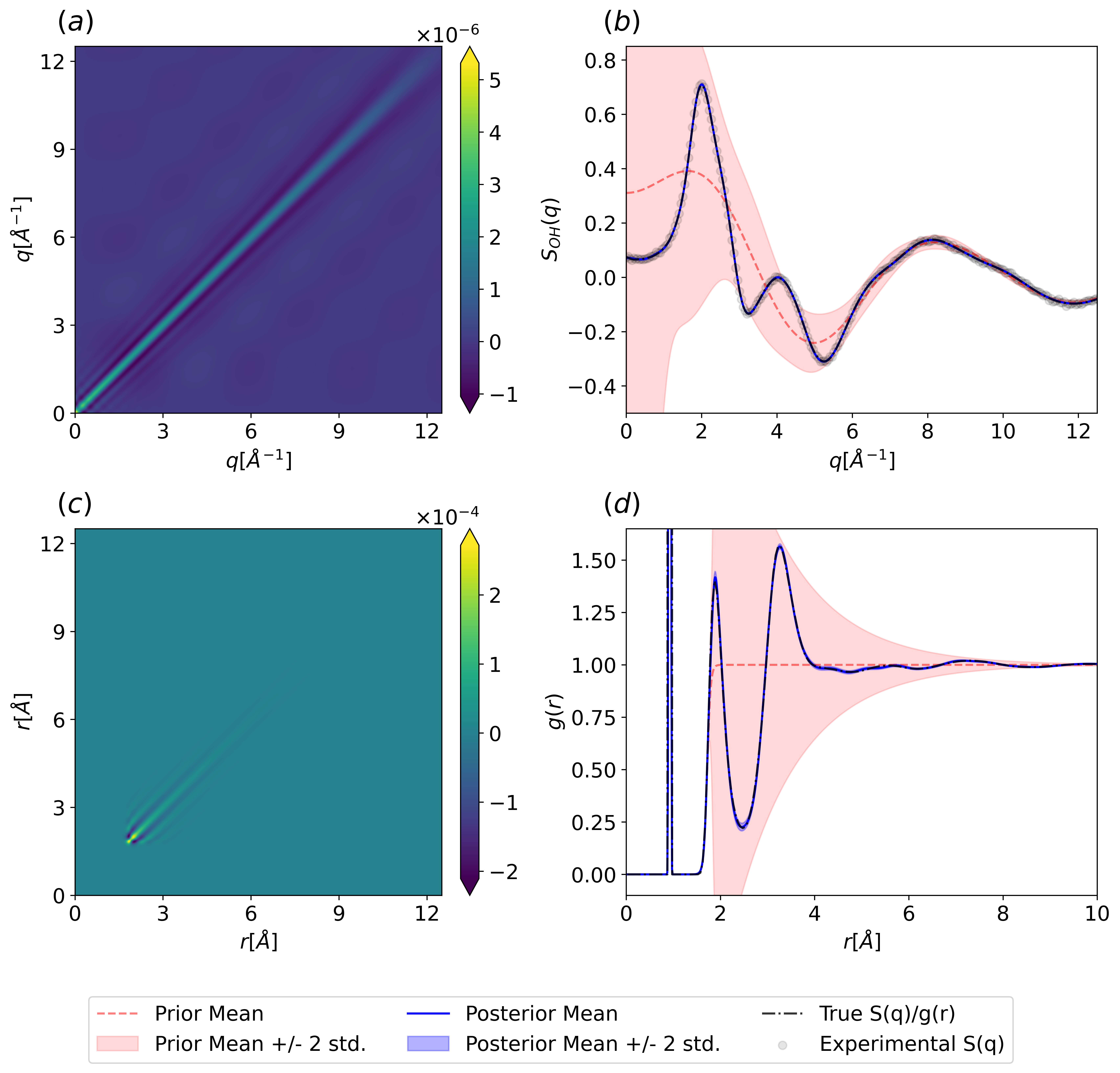}
    \caption{Posterior of the Gaussian process fit to structure factors derived from flexible TIP4P/2005f water for the oxygen-hydrogen correlation. Top left corresponds to eq \eqref{eq:SQ_post_sigma}. Top right corresponds to eq \eqref{eq:RDF_post_sigma}. Bottom left corresponds to eq \eqref{eq:SQ_post_mu} and the diagonal of the covariance shown in the top left. Bottom right corresponds to eq \eqref{eq:RDF_post_mu} and the diagonal of the covariance shown in the top right. The colorbars are clipped similarly to Figure \ref{fig:kernelviz}.}
    \label{fig:sim_water_fig_OH}
\end{figure}

\begin{figure}
    \centering
    \includegraphics[width=1\linewidth]{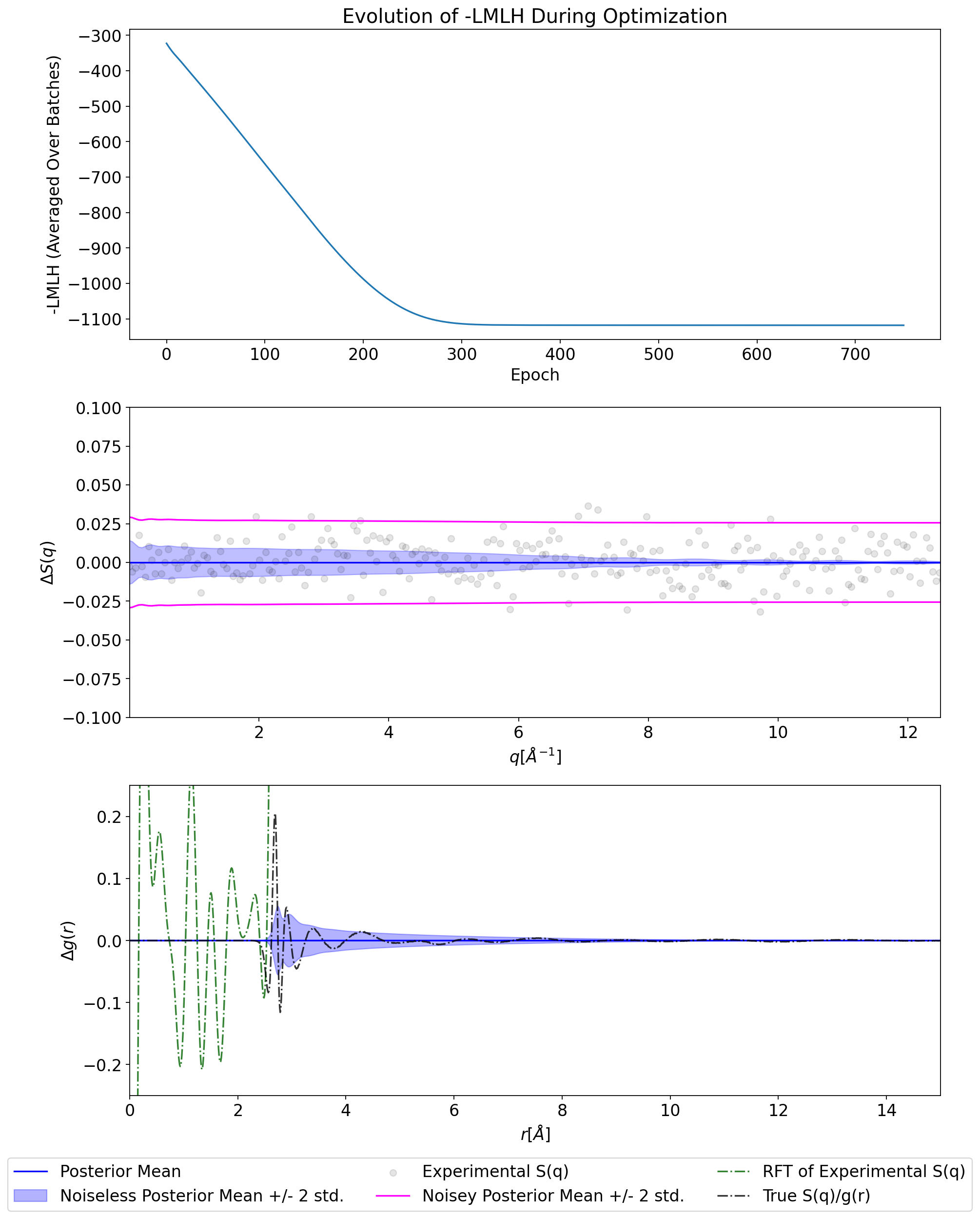}
    \caption{Top: The negative log marginal likelihood as a function of epoch during gradient descent optimization of equation \ref{eq:opt_target}. Middle: Simulated OO structure factor residuals. Bottom: Simulated OO radial distribution function residuals.}
    \label{fig:LMLH_evolution_SimWater_OO}
\end{figure}

\begin{figure}
    \centering
    \includegraphics[width=1\linewidth]{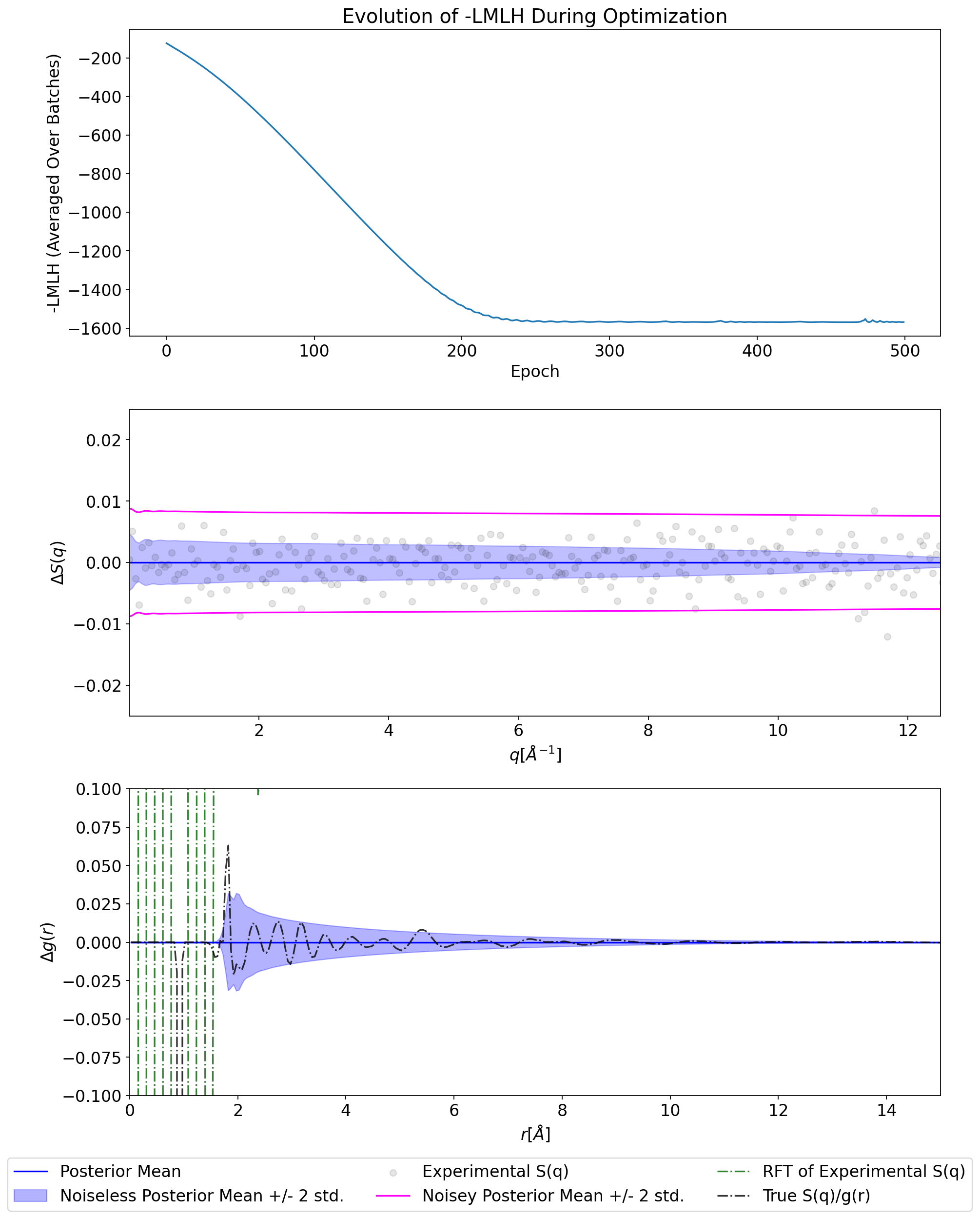}
    \caption{Top: The negative log marginal likelihood as a function of epoch during gradient descent optimization of equation \ref{eq:opt_target}. Middle: Simulated OH structure factor residuals. Bottom: Simulated OH radial distribution function residuals.}
    \label{fig:LMLH_evolution_SimWater_OH}
\end{figure}

\begin{figure}
    \centering
    \includegraphics[width=1\linewidth]{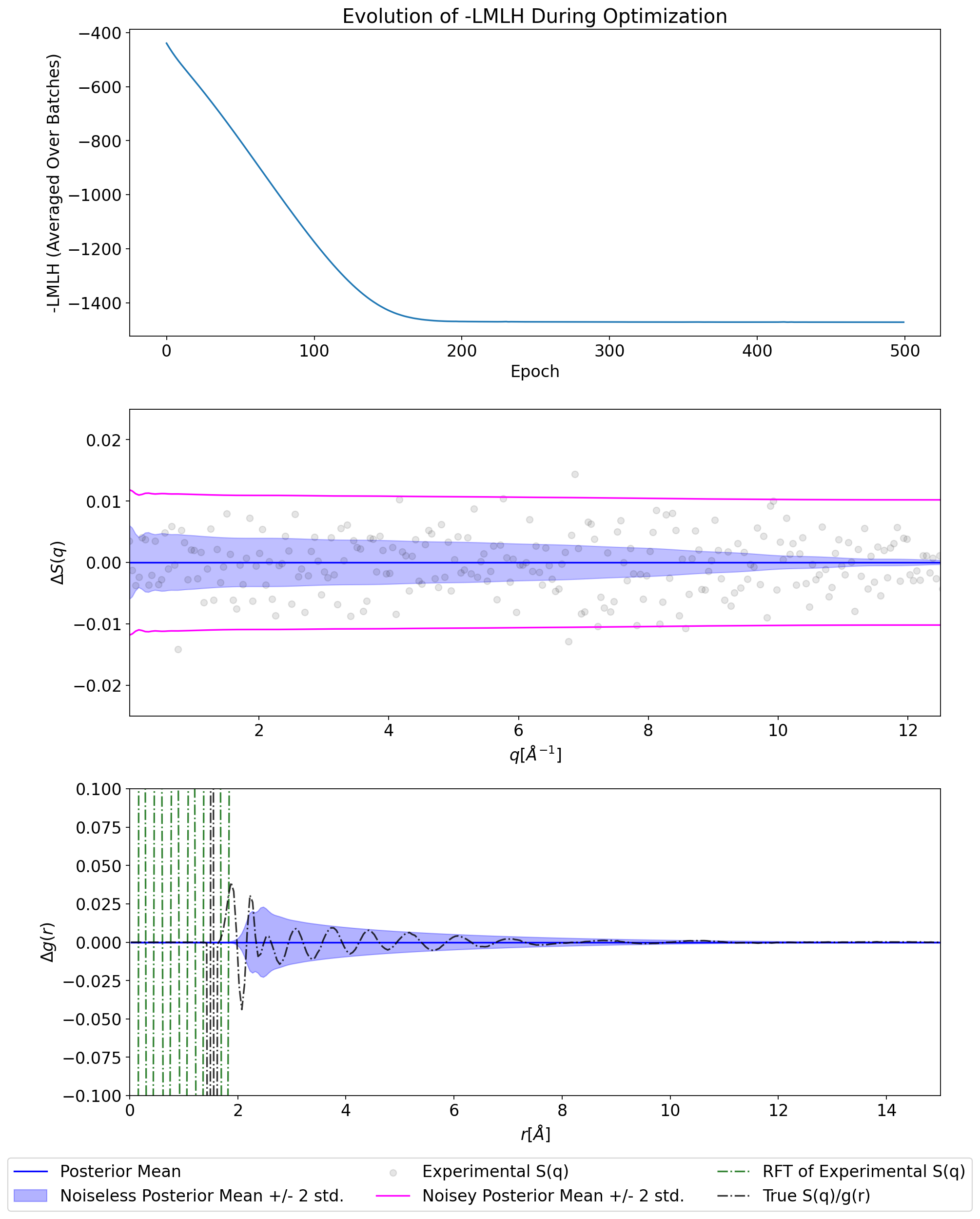}
    \caption{Top: The negative log marginal likelihood as a function of epoch during gradient descent optimization of equation \ref{eq:opt_target}. Middle: Simulated HH structure factor residuals. Bottom: Simulated HH radial distribution function residuals.}
    \label{fig:LMLH_evolution_SimWater_HH}
\end{figure}

\newpage 
\section{Experimental Water Analysis}

Provided below are the log-marginal likelihood evolution and residuals for the oxygen-oxygen partial structure factor from the experimental X-ray scattering dataset. Optimized hyperparameters are provided in Table \ref{tab:init_opt_params_OO_exp}.

\begin{figure}[H]
    \centering
    \includegraphics[width=0.95\linewidth]{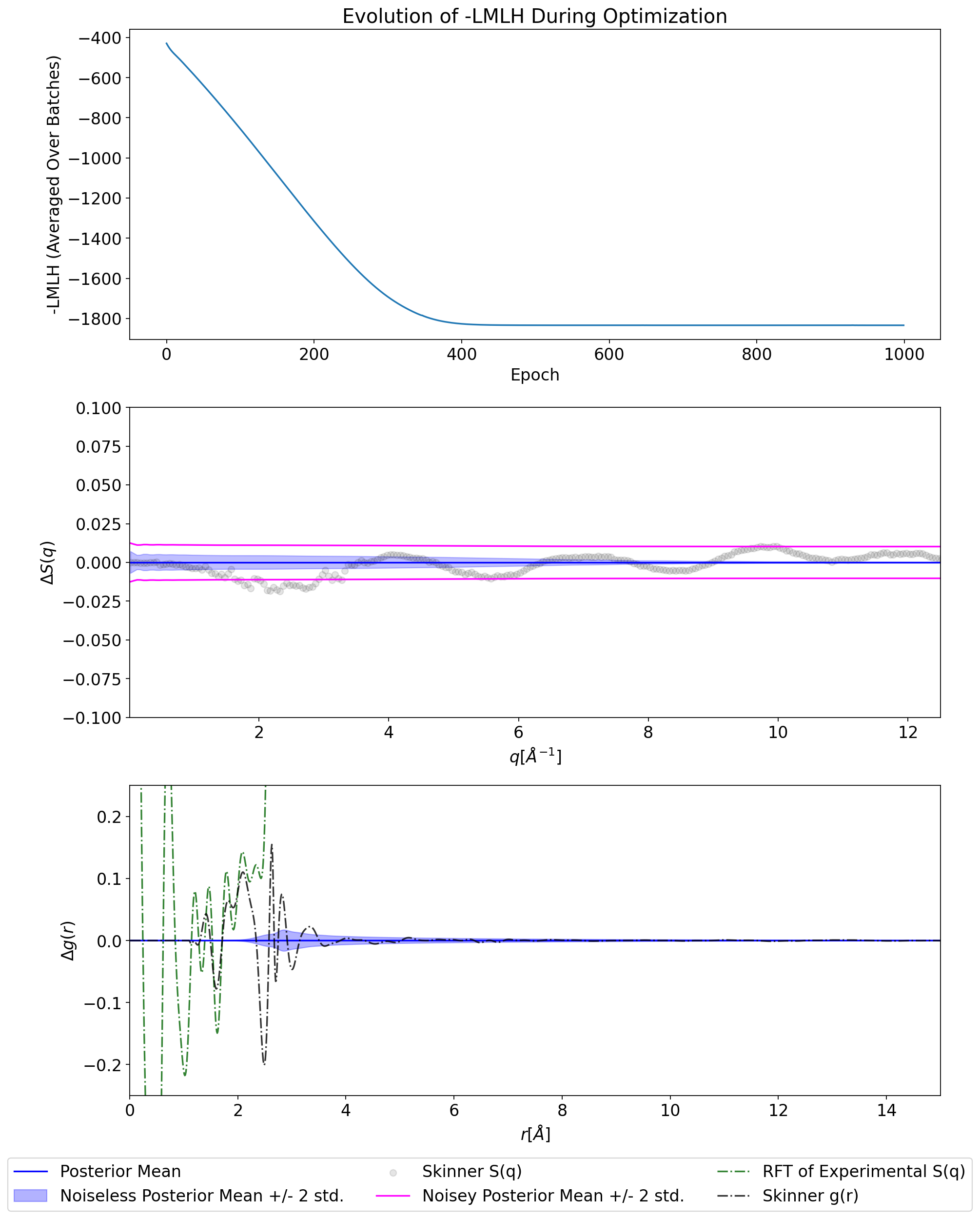}
    \caption{Top: The negative log marginal likelihood as a function of epoch during gradient descent optimization of
equation 39. Middle: OO structure factor residuals. Bottom: OO radial distribution function error comparison between prior (red), posterior (dark blue) and skinners (light blue).}
    \label{fig:OO_SI}
\end{figure}

\begin{table}[H]
    \centering
    \begin{tabular}{|c|c|c|c|c|c|c|}
        \hline
        Parameter & Initial Value & Lower Bound & Upper Bound & Optimized Value & $\Delta$ & \% Change \\
        \hline
        $\ell$       & 1.230821 & 0.100000 & 2.500000 & 0.808131 & $-0.422690$ & $-17.61\%$ \\
        Max          & 3.000000 & 0.200000 & 10.000000 & 7.977926 & \phantom{$-$}4.977926 & \phantom{$-$}50.80\% \\
        Slope        & 5.678962 & 1.000000 & 50.000000 & 11.264938 & \phantom{$-$}5.585976 & \phantom{$-$}11.40\% \\
        Loc          & 2.948104 & 0.500000 & 6.000000 & 2.790531 & $-0.157573$ & $-2.86\%$ \\
        Decay        & 0.534100 & 0.050000 & 3.000000 & 0.895475 & \phantom{$-$}0.361375 & \phantom{$-$}12.25\% \\
        $\omega$   & 0.147792 & 0.000100 & 0.500000 & 0.005117 & $-0.142675$ & $-28.54\%$ \\
        $r_0$        & 2.810502 & 0.010000 & 20.000000 & 2.662684 & $-0.147818$ & $-0.74\%$ \\
        $s$          & 25.084942 & 0.001000 & 50.000000 & 45.946390 & \phantom{$-$}20.861448 & \phantom{$-$}41.72\% \\
        \hline
    \end{tabular}
    \caption{Initial parameters, optimized parameters, and their bounds for experimental OO dataset.}
    \label{tab:init_opt_params_OO_exp}
\end{table}

\printbibliography

\end{document}